\documentclass[review]{elsarticle}

\usepackage[colorlinks=true,linkcolor=blue]{hyperref}%
\usepackage{lineno}
\usepackage{arydshln}
\usepackage{multirow}
\usepackage[normalem]{ulem}
\usepackage{soul}
\modulolinenumbers[5]
\journal{NeuroImage}
\biboptions{semicolon,round,sort,authoryear}

\begin{document}
\begin{frontmatter}
\title{Development of accurate human head models for personalized electromagnetic dosimetry using deep learning}

\author[a,b,c]{Essam A. Rashed}
\ead{essam.rashed@nitech.ac.jp}
\author[a]{Jose Gomez-Tames}
\author[a,d]{Akimasa Hirata}
\address[a]{Department of Electrical and Mechanical Engineering, Nagoya Institute of Technology, Nagoya 466-8555, Japan}
\address[b]{Department of Computer Science, Faculty of Informatics \& Computer Science, \\The British University in Egypt, Cairo 11837, Egypt}
\address[c]{Department of Mathematics, Faculty of Science, Suez Canal University,\\ Ismailia 41522, Egypt}
\address[d]{Center of Biomedical Physics and Information Technology, Nagoya Institute of Technology, Nagoya 466-8555, Japan}

%=============
% Abstract
%=============

\begin{abstract}

The development of personalized human head models from medical images has become an important topic in the electromagnetic dosimetry field, including the optimization of electrostimulation, safety assessments, etc. Human head models are commonly generated via the segmentation of magnetic resonance images into different anatomical tissues. This process is time consuming and requires special experience for segmenting a relatively large number of tissues. Thus, it is challenging to accurately compute the electric field in different specific brain regions. Recently, deep learning has been applied for the segmentation of the human brain. However, most studies have focused on the segmentation of brain tissue only and little attention has been paid to other tissues, which are considerably important for electromagnetic dosimetry.

In this study, we propose a new architecture for a convolutional neural network, named ForkNet, to perform the segmentation of whole human head structures, which is essential for evaluating the electrical field distribution in the brain. The proposed network can be used to generate personalized head models and applied for the evaluation of the electric field in the brain during transcranial magnetic stimulation. Our computational results indicate that the head models generated using the proposed network exhibit strong matching with those created via manual segmentation in an intra-scanner segmentation task.

\end{abstract}

\begin{keyword}
convolutional neural network, deep learning, image segmentation, transcranial magnetic stimulation 
\end{keyword}

\end{frontmatter}

%\linenumbers

%==========================
% 1. Introduction
%==========================

\section{Introduction}

Dramatic progress has been made in electromagnetic (non-ionizing radiation) dosimetry over the last decades \citep{Reilly2016PMB}. One key feature of electromagnetic dosimetry is that the anatomical complexity of biological tissues, which are the physical agents for stimulation, results in complex electric field distributions. In this context, the most remarkable body part is the head. The importance of dose (internal physical quantity) modeling extends to medical applications, human safety from external electromagnetic fields, etc. 

One commonly used application is non-invasive brain stimulation, i.e., techniques such as transcranial direct current stimulation (tDCS) and transcranial magnetic stimulation (TMS), for the purposes of treatment and rehabilitation in neurological clinical applications \citep{Walsh2000nature, Nitsche2008BS}. However, it is challenging to accurately compute the electric field in a specific brain region. Such difficulties are attributable to inter- and intra-subject variability \citep{Laakso2015BS, Lopez2014BS, Maeda2002CN}. Moreover, it is necessary to consider several parameters for optimization, such as those related to the setup of the electrical field source. Namely, coil orientation, position, and design in TMS \citep{Iwahashi2017PMB, GomezTames2018BS, Lee2018CN, Madsen2015BS, Richter2013PO, Deng2013BS} and electrode montage and position in tDCS \citep{Ramaraju2018JNE, Gomez2019JNE, Horvath2014FSN} require special attention. Head models generated via anatomical segmentation can be used to conduct computer simulations to optimize brain stimulation procedures in advance \citep{Holderfer2006CN}. Thus, the generation of personalized head models through the automatic segmentation of the anatomical structures of the head has become essential.

The segmentation of all the tissues that comprise the human head is a time consuming process that requires special skills and long-time experience. Most related works in the literature focus on solving the following two clustering problems: i) identification of the anatomical structures of the brain, namely white matter (WM), gray matter (GM), and cerebrospinal fluid (CSF), and ii) labeling of brain abnormalities, such as tumors. However, to generate a human head model to simulate the induced electrical field, non-brain tissues should also be considered in the segmentation process \citep{Thielscher2011NeuroImage, Lee2018CN, Rashed2019Access}. 

Deep learning is a promising and emerging machine learning technology that has led to remarkable impact in big data analysis and understanding \citep{LeCun1998ProcIEEE, Lecun2015Nature}. Convolutional neural networks (CNNs) are now known as the state-of-the-art technique for image segmentation, especially for highly complicated problems \citep{Ker2018Access}. Deep learning has been used in several brain segmentation tasks to solve the above-mentioned problems \citep{Akkus2017JDI}. Recently, it becomes interesting to investigate how deep learning can contribute to the personalized electromagnetic dosimetry \citep{Rashed2019PULSE}. For anatomical brain structure segmentation, several methods have been presented \citep{Chen2018NeuroImage, Chen2018TMI, Wachinger2018NeuroImage, Mehta2017JMI, Milletari2017CVIU, Moeskops2016TMI, deBrebisson2015CVPR, Zhang2015NeuroImage}. Additionally, methods for the segmentation of small regions within the brain has also been proposed \citep{Roy2019NeuroImage, Roy2019TMI, Dolz2018NeuroImage, Kushibar2018MIA, Choi2016JNM}. Such methods have also been used in brain lesion labeling \citep{Havaei2017MIA, Kamnitsas2017MIA, Pereira2016TMI, Brosch2016TMI}. Moreover, CNNs have been used to segment the brain from other non-brain structures \citep{Mohseni2017TMI}.  

Anatomical brain segmentation was achieved with high accuracy in the above-mentioned studies. In most cases, e.g. VoxResNet \citep{Chen2018NeuroImage}, DRINet \citep{Chen2018TMI}, and Auto-Net \citep{Mohseni2017TMI}, the number of tissues considered was limited to 3-5 only. However, the problem of constructing a model of the whole human head is more complicated. In this study, we consider the annotation of 13 different head tissues/fluids, which requires a more sophisticated design of the network architecture employed. Automatic segmentation to construct accurate human head models is a challenging problem owing to the variations in morphology of several head components. 

In this paper, we propose a new architecture of CNN for the segmentation of human head structures. The proposed architecture was inspired by the well-known U-net architecture \citep{Ronneberger2015MICCAI}, and it consists of several U-net structures with unified encoders and fragment decoders corresponding to different anatomical structures. The network encoders used are fully connected to emphasize the individual segmentation of image features at different levels. Then, decoders are used to identify the anatomical features of the head independently. Via this approach, the proposed architecture can automatically generate head models with high-quality when applied to intra-scanner segmentation where the MRI scan settings of the training and test data match. The constructed head models were evaluated via TMS simulation experiments in terms of the electric field induced in the brain. To the best of our knowledge, this is the first study to propose the construction of a whole human head model using deep learning.

%==========================
% 2. Human head models
%==========================

\section{Human head models}

In the existing literature, several researchers were able to improve the human head model from a simple isotropic sphere to a realistic one. However, the development of realistic models is still a challenging problem with an associated high computational cost and difficulties related to small size and/or low-contrast tissues. Inaccurate segmentation may lead to a false estimation of the electric field distribution, resulting in difficulties for stimulation planning. Moreover, it is important to complete the modeling process in a short time, especially for neurosurgery planning procedures \citep{Picht2009CN}.

Early attempts to develop head models from five types of tissue (skin, skull, CSF, WM, and GM) demonstrated the need for realistic head models rather than standard spheres \citep{Wanger2004TBE}. Later on, more accurate models representing the same five tissues were proposed by \cite{Chen2009JNM}. Segmentation was conducted by performing threshold operations on pixel (voxel) intensity values. Recently, automatic segmentation has been increasingly associated with high-quality brain segmentation software, such as FreeSurfer \citep{Fischl2012NeuroImage}. Another example of automatic segmentation of head tissues using T2-weighted magnetic resonance images (MRI) is presented in \cite{Fortunati2015PMB}. In this work, a fusion of atlas-based and intensity-based segmentation is used to identify eight different tissues; cerebrum, cerebellum, brainstem, myelum, CSF, vitreous humor, sclera, and eye lens. Although the use of automatic segmentation methods can speed up the development of head models, confirmation and fine-tuning are still needed. A useful review of the development of human models can be found in the work of \cite{Kainz2019TRPMS}. This paper presents a deep learning approach for the automatic construction of accurate human head model from T1-weighted MRI data.

%==========================
% Figure-1
%==========================

\begin{figure*}
\centering
\includegraphics[width=\textwidth]{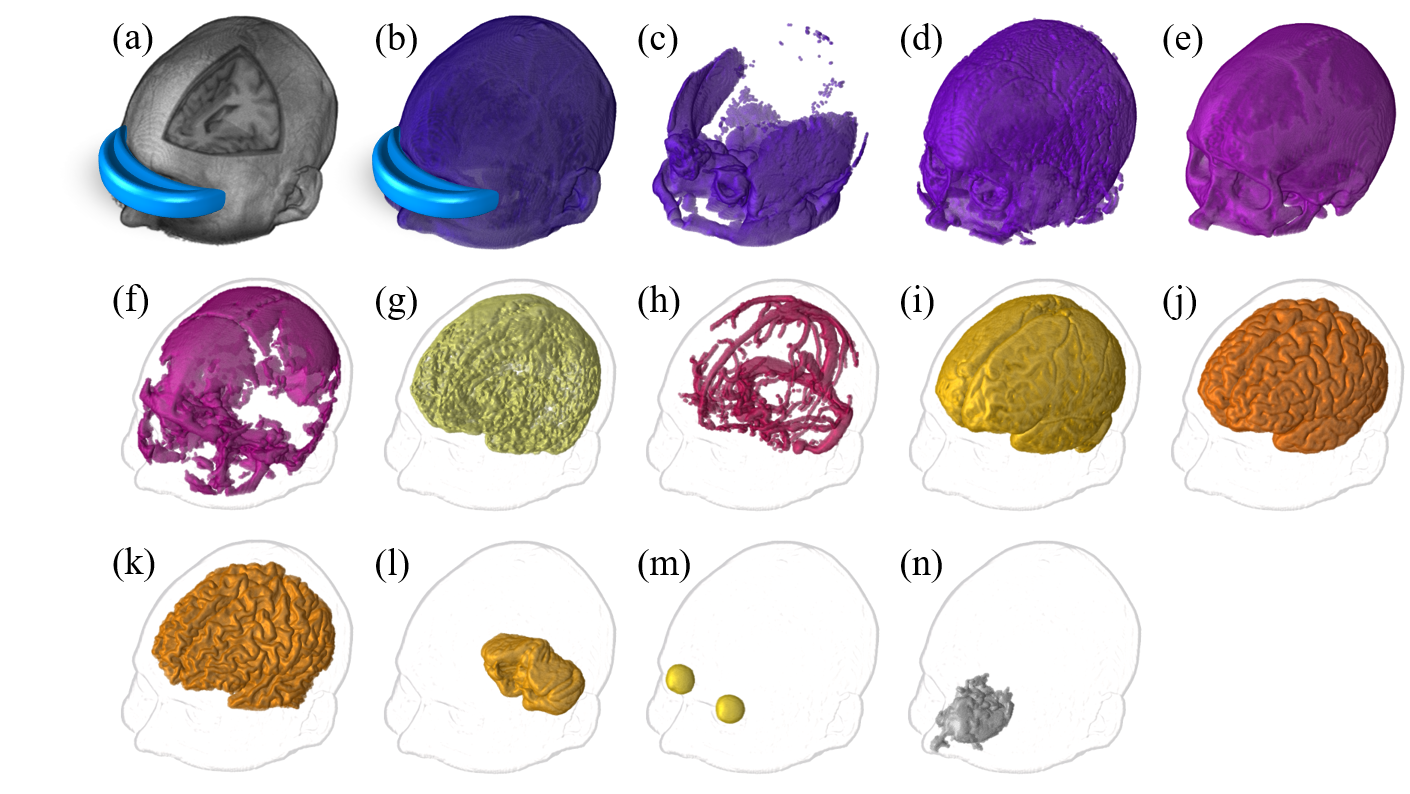}
\caption{Example of the construction of a human head model via segmentation using (a) 3.0T MRI acquisition to generate (b) skin, (c) muscle, (d) fat, (e) bone (cortical), (f) bone (cancellous), (g) dura, (h) blood vessels, (i) CSF, (j) GM, (k) WM, (l) cerebellum, (m) vitreous humor, and (n) mucous tissue.}
\label{seg}
\end{figure*}

%==========================
% 3. Materials and Methods
%==========================

\section{Materials and methods}

%==========================
% 3.1 MRI dadaset
%==========================

\subsection{MRI dataset}

Structural MRI scans with a voxel size of 1.0$\times$1.0$\times$1.0 mm of 18 subjects were obtained from a freely available dataset (NAMIC: Brain Multimodality). Images are acquired using 3T GE scanner at BWH in Boston, MA with field of view 256 cm$^2$ and slice matrix of 256$\times$256 pixels. More details about imaging setup can be found in NAMIC\footnote{ \href{http://hdl.handle.net/1926/1687} {http://hdl.handle.net/1926/1687}}. All the images were obtained from male subjects (in the age range of 43.44 $\pm$ 9.77 years). A semi-automatic segmentation procedure was used to generate head models for all the subjects (hereinafter referred to as original head models $R^\circ$). The data from each subject was segmented into skin, muscle, fat, bone (cortical), bone (cancellous), dura, blood vessels, CSF, GM, WM, cerebellum, vitreous humor, and mucous tissue. An example of segmentation corresponding to (case01035) is shown in Fig.~\ref{seg}. The semi-automatic method used a registered T1- and T2-weighted MRI images to segment head tissues using region-growing and thresholding techniques. Several structural-based constrains are set to improve the segmentation accuracy and manual correction is also used. Therefore, the segmentation accuracy of the semi-automatic method has some limitations especially for non-brain tissues. Details of the segmentation techniques used in this study can be found in the work of \cite{Laakso2015BS}. 

Both MRI images (T1 only) and the segmented models were set to $256^3$ voxels. MRI data were normalized with zero mean and unit variance, followed by scaling in the range of $[0,1]$ without additional contrast enhancement corrections, and each voxel value was assigned a storage size of 32 bits (floating point). All images are used without performing mutual registration between different subjects and we did not include additional augmented data. 

%======================
% Figure-2
%======================

\begin{figure*}
\includegraphics[width=\textwidth]{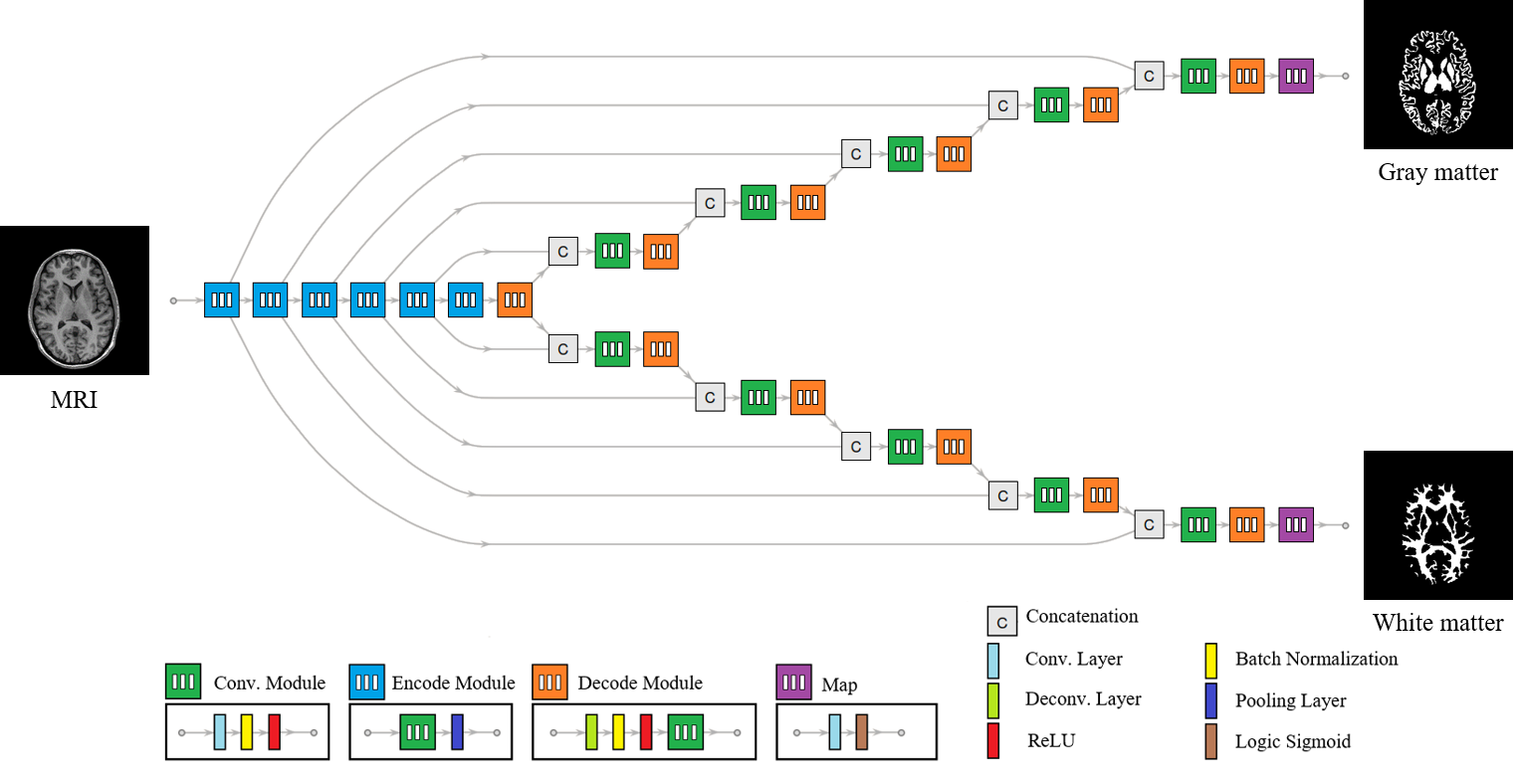}
\caption{Detailed architecture of ForkNet with a degree of $N=2$ with colored layer identification keys. An example of an input MRI image and output binary labels for GM and WM are shown. The computation of the feature variables for each layer is demonstrated in detail in Table~\ref{ForkSize}.}
\label{ForkNet}
\end{figure*}

%==========================
% 3.2 Network Architecture
%==========================

\subsection{Network architecture}

The following two approaches are commonly used for the development of deep CNN architectures for brain segmentation problems: i) patch-based CNNs, in which an image is divided into patches to extract pixel-oriented features considering the local neighborhood, and ii) semantic-based CNN, in which the whole image is used as a single training datum. The former architecture has an advantage in terms of the availability of relatively large training sets because a single image can be split into several patches with variable overlaps. However, patch-based CNNs may lose global information owing to atypical partition schemes. The latter architecture reduces the burden of image pre-processing. Nevertheless, it requires a relatively large training set to achieve high accuracy. In this study, a semantic-based CNN architecture was developed, in which end-to-end mapping is conducted for the whole image.

The proposed CNN architecture was designed to connect a single input (i.e., an MRI image) with multiple ($n=1,\dots,N$) outputs (i.e., the segmentation labels of different anatomical structures). The proposed network is shown in Fig.~\ref{ForkNet}, and the feature size for each layer is shown in detail in Table~\ref{ForkSize}. The proposed architecture, named ForkNet, comprises 23 layers. All convolutions are conducted with a kernel size of 3$\times$3, a stride of 1$\times$1, and a padding of 1. All deconvolutions are conducted with a kernel size of 2$\times$2, and max pooling is conducted with a kernel size of 2$\times$2. The batch normalization layer is computed with a momentum of 0.9 and a stability parameter $\epsilon$ = 0.001. The input is a 2D slice (256$\times$256 pixels) extracted from the MRI images and the $n^{th}$ output is the corresponding 2D (256$\times$256 pixels) binary label image.

%======================
% Table-1
%====================== 

\begin{table*}
\centering
\footnotesize
\label{ForkNetTab}
\caption{Detailed architecture of ForkNet (shown in Fig.~\ref{ForkNet}) with degree $N$.}
\begin{tabular}{|l|lllll|}
\hline 
{\bf Module} & {\bf Layer} & {\bf Output size} & {\bf Kernel} & {\bf Stride} & {\bf Padding}\\
\hline \hline
Input & & $2^{8}\times2^{8}$ &&&\\
\hline
EncMod$_{i}$ & Convolution & $2^{(i+2)} \times 2^{(9-i)} \times 2^{(9-i)}$ & ($2^{(i+2)}$ $\times$3$\times$3) & (1$\times$1) & (1$\times$1)\\
$i=1 \rightarrow 6$& BN \& ReLU & $2^{(i+2)} \times 2^{(9-i)} \times 2^{(9-i)}$ & &  & \\
& Pooling (Max) & $2^{(i+2)} \times 2^{(8-i)} \times 2^{(8-i)}$ &&& \\
\hline
DecMod$_{j,n}$ & Deconvolution & $2^{(j+1)} \times 2^{(9-j)} \times 2^{(9-j)}$ & ($2^{(j+1)}$ $\times$2$\times$2) & (2$\times$2) & \\
$j=6 \rightarrow 1$ & BN \& ReLU & $2^{(j+1)} \times 2^{(9-j)} \times 2^{(9-j)}$ & &  & \\
 & Convolution & $2^{(j+1)} \times 2^{(9-j)} \times 2^{(9-j)}$ & ($2^{(j+1)}$ $\times$3$\times$3) & (1$\times$1) & (1$\times$1)\\
\hline
ConvMod$_{j,n}$ & Convolution & $2^{(j+2)} \times 2^{(8-j)} \times 2^{(8-j)}$ & ($2^{(j+2)}$ $\times$3$\times$3) & (1$\times$1) & (1$\times$1)\\
$j=5 \rightarrow 1$ & BN \& ReLU & $2^{(j+2)} \times 2^{(8-j)} \times 2^{(8-j)}$ & &  & \\
\hline
Concat$_{j,n}$ & Concatenation &  $2^{(j+3)} \times 2^{(8-j)} \times 2^{(8-j)}$ & &&\\
$j=5 \rightarrow 1$ &  &  & &&\\
\hline
Map$_n$ & Convolution & $1 \times 2^{8} \times 2^{8}$ & ($1\times3\times3$)&($1\times1 $)&($1\times1$)\\
$n=1 \rightarrow N$  & Sigmoid (Log) & $2^{8} \times 2^{8}$ & &&\\
\hline
Output$_n$ & & $ 2^{8}\times2^{8}$ &&&\\
$n=1 \rightarrow N$ & &  &&&\\
\hline
%}
\end{tabular}
\label{ForkSize}
\end{table*}

The main feature of the proposed architecture is that the encoder layers comprise a single track whereas the decoder layers are positioned in parallel. This novel design has several advantages over conventional CNN architectures. For instance, i) this design can handle the problem of end-to-end semantic network confusion that occurs when the number of segmentation labels is relatively large (here, $N=13$). Each decoder terminal leads to a binary image that corresponds to a single tissue label. Moreover, ii) split decoder provide the ability to perform segmentation with different resolutions within the same network. For example, it is possible to change the number of decoder module layers to obtain higher or lower resolutions for specific anatomical structures. Multi-resolution ForkNet allows for learning using anatomical structures with different resolutions in a single step. Of course, the design of multi-resolution architecture should precede the training phase. Additionally, iii) it is possible to customize the design of each decoder independently, which may be more suitable owing to the texture variability of different anatomical structures. Therefore, the proposed architecture provides a more general design that can be adapted to the standard U-net architecture by setting $N=1$.

Considering a 3D MRI image with $K$ slices, the output of the proposed network is computed as:

% EQ-1
\begin{equation}
L_{k,n}=\textnormal{ForkNet}(M_{k}), ~k=1,\dots,K,~ n=1,\dots,N,
\end{equation} 
where $M_k$ is an MRI slice and $L_{k,n}$ is the network output with elements $L_{k,n}(i,j) \in [0,1]$. The network-generated head model is computed using the SoftMax rule:

%EQ-2
\begin{equation}
R_{k}(i,j)=\arg \max_{n} L_{k,n}(i,j), \forall~~ i,j,~k=1,\dots,K.
\end{equation} 
In other words, within a single pixel, we set the label $n$ with the maximum corresponding network output value.

%======================
% Figure-3
%======================

\begin{figure*}
\center
\includegraphics[width=.95\textwidth]{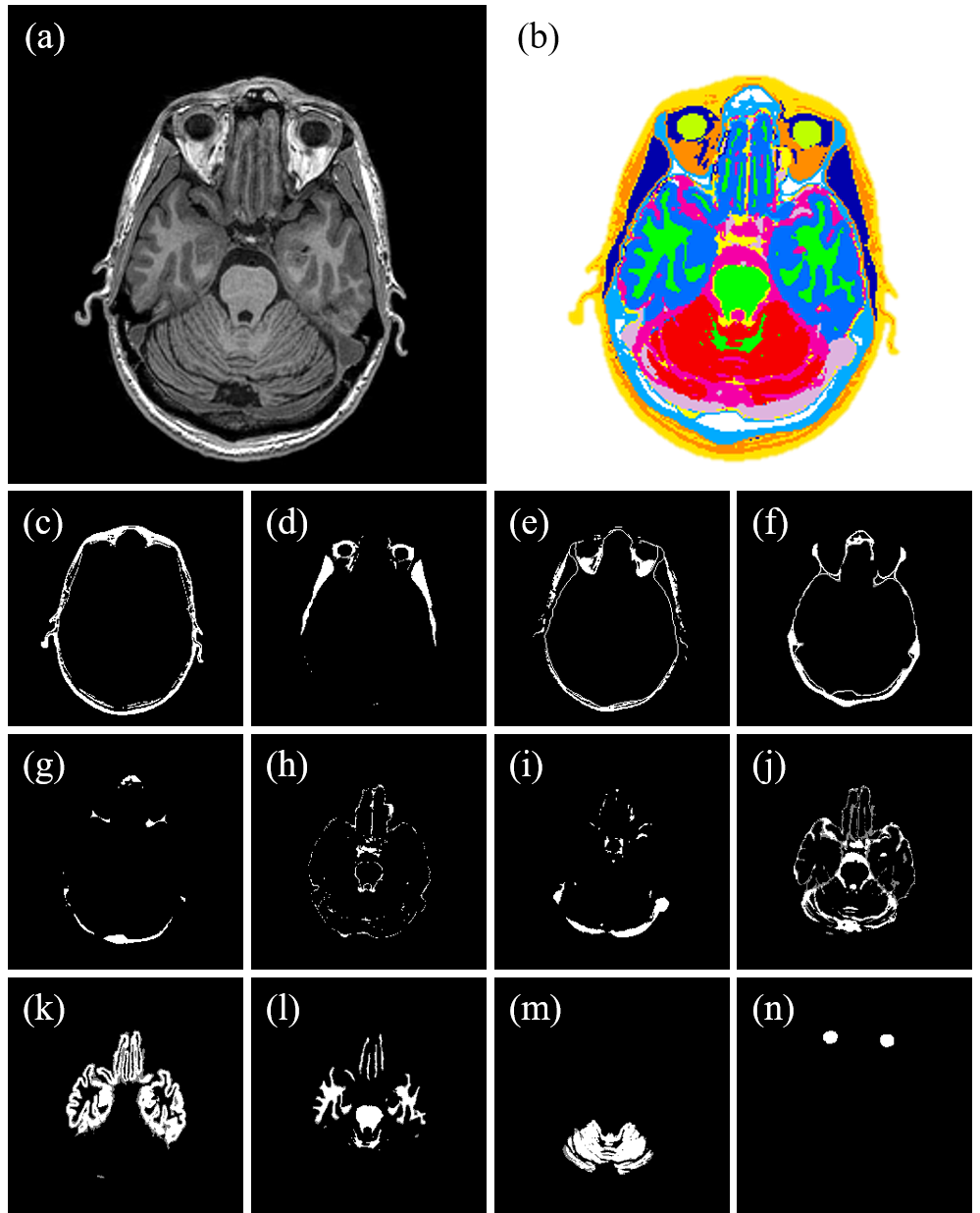}
\caption{(a) Sample T1 MRI axial slice with its corresponding segmented head model, shown in (b). Binary images used to train different tissues representing (c) skin, (d) muscle, (e) fat, (f) bone (cortical), (g) bone (cancellous), (h) dura, (i) blood vessels, (j) CSF, (k) GM, (l) WM, (m) cerebellum, and (n) vitreous humor.}
\label{P5Seg}
\end{figure*}

%==========================
% 3.3 Training strategy
%==========================

\subsection{Training strategy}

In all training phases, we applied a cross-validation method in which an arbitrary subject is excluded from the dataset and the remaining subjects (17 volumes, except when the use of a different setup is stated) are used for the current training phase. The cross-entropy loss function is minimized using the ADAM algorithm \citep{Kingma2014arXiv}. We considered the automatic computed learning rate\footnote{Computed using Wolfram (R) Mathematica (R) function \texttt{NetTrain}} and a batch size of 2. The MR slices of training data are randomly shuffled, and then 90\% of shuffled slices are used for training and 10\% are used for validation to avoid training over-fitting. A sample of images used for a training instance is shown in Fig.~\ref{P5Seg}. For all the results presented in this study, we assume that all data are extracted using an identical decoder structure, and all labels are defined with a unified resolution. Therefore, unified encoders are trained to extract the features of the MRI data at different resolution levels. These features are concatenated to decoders at corresponding levels (Fig.~\ref{ForkNet}). However, the decoders are mostly oriented to a single anatomical structure at each network branch. We found that this strategy emphasizes the individual segmentation of image features at different levels, leading to higher segmentation accuracy.

%======================
% Figure-4
%======================

\begin{figure*}
\includegraphics[width=\textwidth]{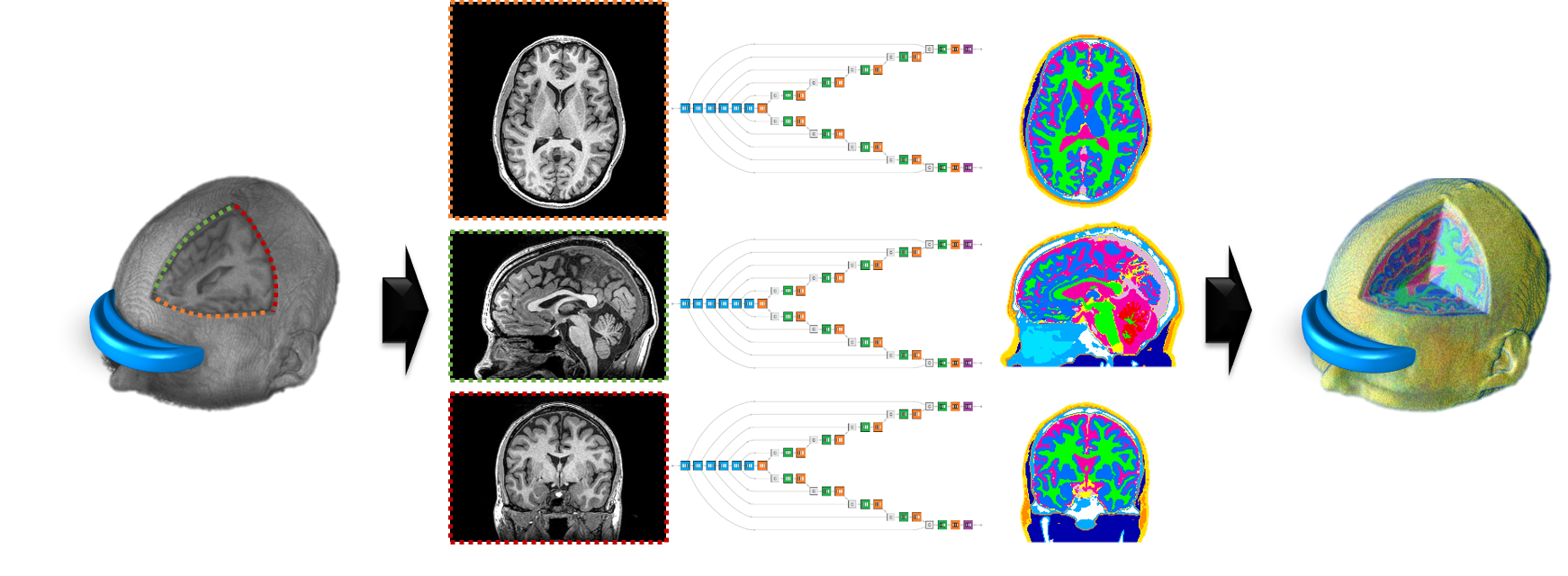}
\caption{Generation of a head model using different slicing schemes: axial (top), sagittal (middle), and coronal (bottom). The results obtained from three networks are aggregated to generate the final head model.}
\label{ForkNet3D}
% /home/essam/FreeModelresults/J_L17_P04_50_2/Orig_no90s_P04_print.raw
\end{figure*}

%============================
% 3.4 Head model formulation
%============================

\subsection{Head model formulation}

In an early stage of this work, we found that using axial slices can lead to good segmentation quality within the slice space. However, this is not enough for head model formulation. In neuroimages, it is important to ensure segmentation quality in 3D, especially for anatomical structures that appear with low contrast and/or limited regions in MRI images. It was difficult to directly extend ForkNet to 3D owing to memory limitations in the available computing facilities. Instead, we considered a 2.5D approach by training individual networks in different slicing directions. The network is trained using three slicing directions (i.e., the axial ($\alpha$), sagittal ($\beta$), and coronal ($\gamma$)), as shown in Fig.~\ref{ForkNet3D}. The final head model $R^{\psi}$ is computed using the following rule:

%EQ-3
\begin{equation}
R^{\psi}_k(i,j) = \left\{
\begin{array}{@{}l@{\thinspace}l}
R^{\alpha}_k(i,j)&: R^{\alpha}_k(i,j) = R^{\beta}_k(i,j) =R^{\gamma}_k(i,j) \\
R^{\alpha}_k(i,j)  &:  R^{\alpha}_k(i,j) = R^{\beta}_k(i,j) \neq  R^{\gamma}_k(i,j)\\
R^{\beta}_k(i,j)  &:  R^{\alpha}_k(i,j) \neq R^{\beta}_k(i,j) =  R^{\gamma}_k(i,j)\\
R^{\alpha}_k(i,j)  &:  R^{\alpha}_k(i,j) = R^{\gamma}_k(i,j) \neq  R^{\beta}_k(i,j)\\
\textnormal{Fuzzy}  &: R^{\alpha}_k(i,j) \neq R^{\beta}_k(i,j) \neq  R^{\gamma}_k(i,j)\\
\end{array},
\right.
\label{eq03}
\end{equation}
where $R^\alpha$, $R^\beta$, and $R^\gamma$ are the head models computed from the axial, sagittal, and coronal directions, respectively. The rule-based segmentation merge approach in  Eq.~(\ref{eq03}) is voxel-wise a majority vote. When no majority in a voxel is found, additional computations are required. A potential approach (used here) is a neighborhood majority vote using the following equation:

%EQ-4
\begin{equation}
R^{\psi}_k(i,j)= \arg \max_{t=\alpha, \beta, \gamma} \max_n \textnormal{Count}_{i,j \in \Omega} R_k^t(i,j),
\end{equation}
where $\Omega$ is the neighborhood region (here, we used $\Omega$ = 3$\times$3 window). In other words, the mismatched voxels are decided according to the most frequent labels located in their vicinity along the three slicing directions. As shown below in Section 4.1, the percentage of fuzzy voxels is generally very small and it is presented in sparse form. However, it is still possible to extend the window $\Omega$ to a larger size if condensed (non-sparse) fuzzy voxels are observed. Alternatively, one can assign higher priority to some direction (e.g. axial) as the results presented in Section 4.1 indicate that $R^\alpha$ is of slightly higher quality compared $R^\beta$ and $R^\gamma$.

%======================
% Figure-5
%======================

\begin{figure*}
\center
\includegraphics[width=\textwidth]{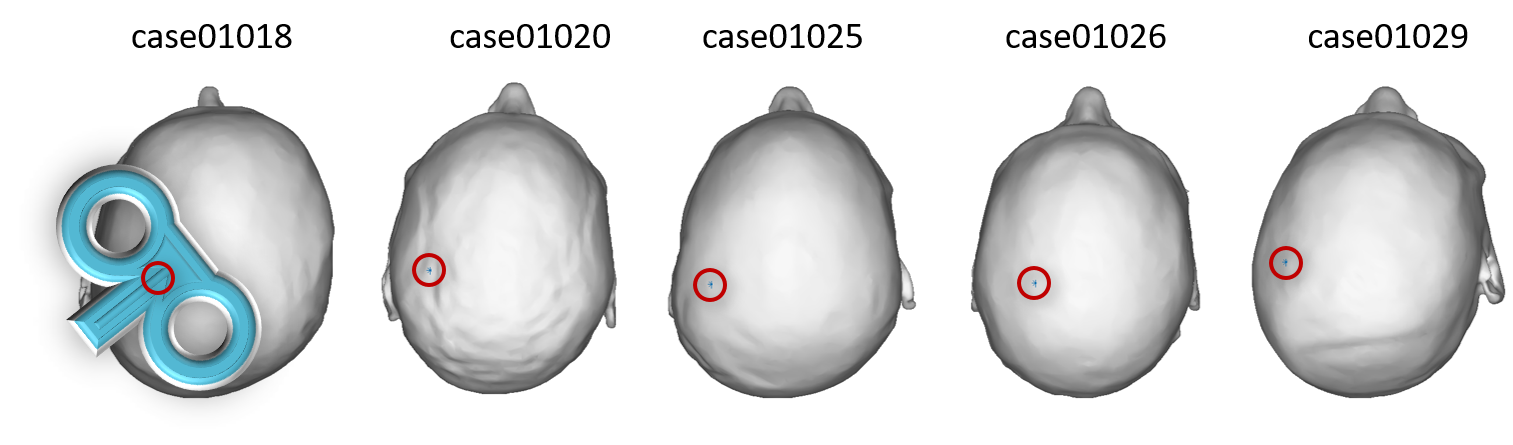}
\caption{TMS coil position setup for brain stimulation targeting the hand motor cortex of different subjects.}
\label{TMSpos}
\end{figure*}

%===============================
% 3.5 Electric Field Simulation
%===============================

\subsection{Electric field simulation}

The head model is used to compute the induced electric fields during TMS via physical simulations. The induced electric field is determined from the vector potential using the finite-element method with cubic elements and piece-wise linear basis functions \citep{Laakso2012PMB}. The magnetic vector potential $A_0$ is computed given the scalar potential
%EQ-5
\begin{equation}
 \nabla \left[ \sigma \left( -\nabla \phi - j \omega A_0 \right) \right]=0,
\end{equation}
where $\sigma$ and $\omega$ are the tissue conductivity and angular frequency, respectively. The electric conductivity of the head model tissues is assumed to be linear and isotropic. The tissue conductivity values are computed using a fourth-order Cole--Cole model with a frequency of 10~kHz, as reported in a previous study \citep{Gabriel1996PMB}. The conductivity values were chosen at 10~kHz in Table~\ref{Tab2}, as they are experimentally accurate \citep{Wake2016PMB} and close to the operation frequency of TMS device \citep{Nieminen2015BS} Selection of another typical conductivity values in TMS studies do not affect the current results \citep{GomezTames2018BS} significantly. In the case of the anisotropy of the WM, its effect is weak in the superficial GM \citep{Opitz2011neuroimage} and omitted in this work. A figure-eight magnetic stimulation coil with outer and inner wing diameters of 9.7~cm and 4.7~cm, respectively, is modeled using the thin-wire approximation, and the magnetic vector potential is calculated using the Biot--Savart law. The coil is placed over the scalp to target the hand motor area of the brain, as shown in Fig.~\ref{TMSpos}.

%=================
% TABLE-2
%=================

\begin{table}
\centering
\footnotesize
\caption{Human tissue conductivity values [S/m] and color labels.}
\label{Tab2}
\setlength{\tabcolsep}{3pt}
\begin{tabular}{ |l c c| l c c| }
\hline
{\bf Tissue}	& {\bf Conductivity} & {\bf  Color label}  & {\bf Tissue}	& {\bf Conductivity}	& {\bf  Color label}  \\
\hline
\hline
Blood & 0.70 & \includegraphics[width=.3cm]{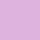}& GM & 0.10&\includegraphics[width=.3cm]{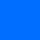}\\
Bone (Canc.) & 0.08 &\includegraphics[width=.3cm]{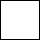}& Mucous tissue& 0.07 &\includegraphics[width=.3cm]{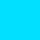}\\
Bone (Cort.) & 0.02 &\includegraphics[width=.3cm]{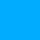}& Muscle &0.34 & \includegraphics[width=.3cm]{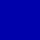}\\ Cerebellum &0.13&\includegraphics[width=.3cm]{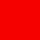}& Skin & 0.10  & \includegraphics[width=.3cm]{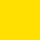}\\
CSF & 2.00&\includegraphics[width=.3cm]{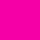} &V.  Humor & 1.50&\includegraphics[width=.3cm]{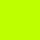}\\
Dura &  0.5 & \includegraphics[width=.3cm]{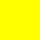}&WM  & 0.07&\includegraphics[width=.3cm]{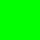}\\
Fat &0.04 &\includegraphics[width=.3cm]{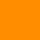}& &&\\
\hline
\end{tabular}
\end{table}

%===============================
% 3.6 Electric Field Simulation
%===============================

\subsection{Data analysis}

To evaluate the segmentation accuracy via measured spatial overlap, the Dice coefficient (DC) is measured and is defined as follows:
%EQ-6
\begin{equation}
DC(A,B)=\frac{2|A \cap B|}{|A|+|B|}.100\%
\end{equation}
where $A$ is a segmentation result and $B$ is the true reference. Also, Hausdorff distance (HD) is used to evaluate segmentation quality as is defined as: 

%Eq-7
\begin{equation}
HD(A,B)= \max_{a\in A} \left[ \min_{b \in B} \left[ d(a,b)\right]\right]
\end{equation}
where $d(.,.)$ is the Euclidian distance between two voxels $a$ and $b$. In TMS, the electric field variation is quantified using the mean absolute error (MAE) error defined as:

%EQ-8
\begin{equation}
\textnormal{MAE}(R^t)=100* \frac{1}{N_c}  \sum_c | EF(R^{\circ}_c)- EF(R^{t}_c)|,
\end{equation}
where $c$ identifies the hand motor area with $N_c$ voxels. Moreover, the MAE values over the hotspot are computed by setting $c$ to the voxels with $EF(R^{\circ}_c)>0.7 \max(EF)$.

%======================
% Figure-6
%======================

\begin{figure*}
\center
\includegraphics[width=\textwidth]{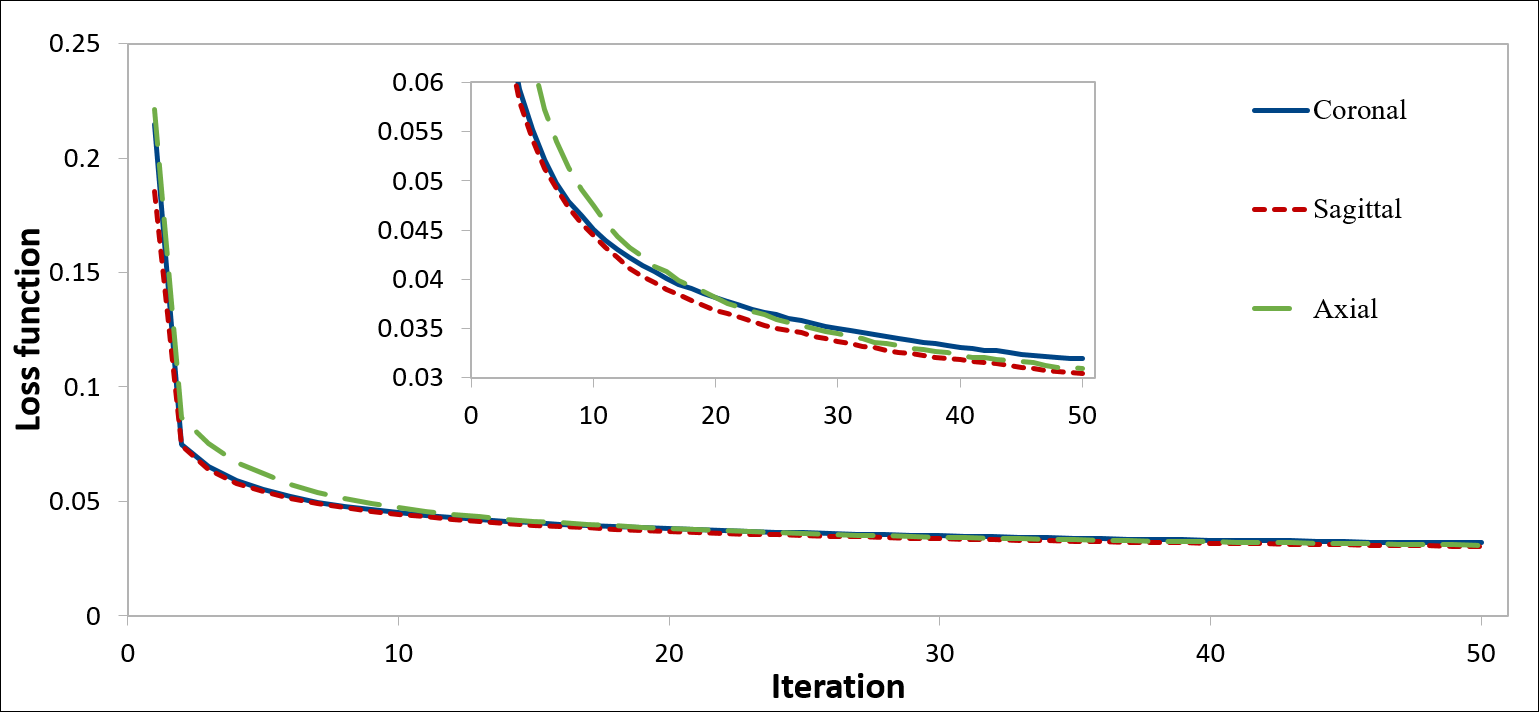}
\caption{Loss functions computed during training using three slicing directions. It can be observed that after few iterations (approximately, 20), the loss function values are almost identical.}
\label{P05_Loss}
\end{figure*}

%==========================
% 4. Results
%==========================

\section{Experimental studies}

%==========================
% 4.1 Head model construction
%==========================

\subsection{Head model construction}

\begin{figure*}
\center
\includegraphics[width=\textwidth]{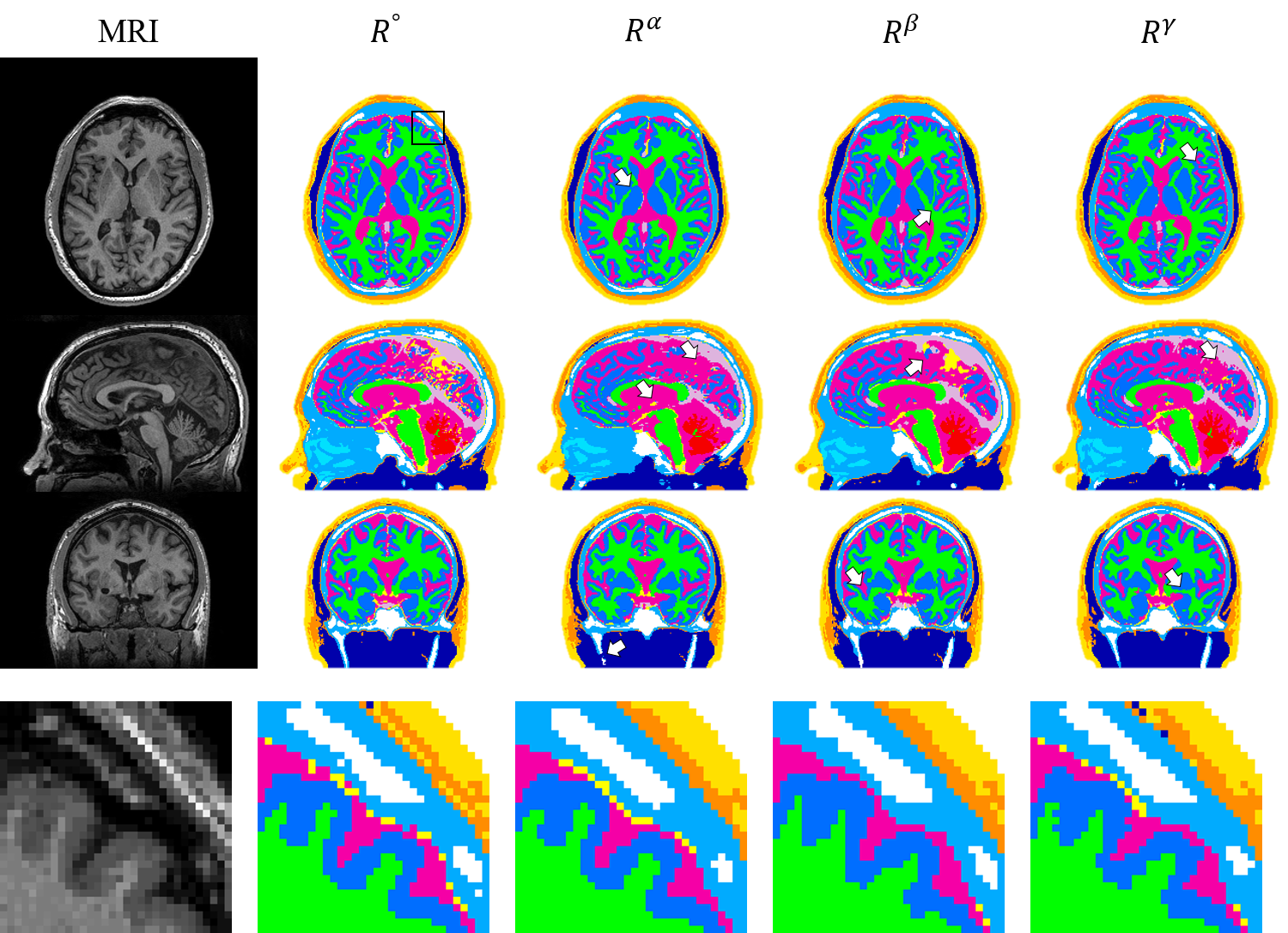}
\caption{From left to right, MRI, original head model $R^\circ$, and network-generated head models extracted from the axial $R^\alpha$, sagittal $R^\beta$, and coronal $R^\gamma$ directions for (case01020). From top to bottom, the central axial, sagittal, and coronal slices. The region labeled with a square in central axial $R^\circ$ is magnified (with corresponding regions in other models) in the last row for clarification. Arrows indicate some regions where mismatching is observed.}
\label{P05}
% /home/essam/FreeModelresults/J_L17_P05_50_2/Model_T_print.raw
\end{figure*}

The proposed architecture, ForkNet, was implemented using Wolfram Mathematica\footnote{Wolfram Research, Inc., Champaign, IL, 2018} (R) Ver. 11.3 , installed on a workstation of 4$\times$Intel (R) Xeon CPUs @ 3.60 GHz, 128 GB of memory and 3$\times$NVIDIA GeForce GTX 1080 GPUs. Computations were conducted using GPUs to speed up the training process. A randomly selected subject (case01020), who was a 53-year-old male, was excluded from the dataset for testing. The remaining subjects were used for training over 50 iterations. The training phase for each single slicing direction was completed in approximately 43 minutes. The loss functions for the three slicing directions are shown in Fig.~\ref{P05_Loss}. From this figure, it can be seen that the behavior of the loss functions was consistent independently of the slicing direction used. The MRI of the subject under study was then evaluated using the trained network. The segmentation results are shown in Fig.~\ref{P05}. The percentage of head voxels that were matched in all three slicing directions was 85.78\%, that for voxels with matching values in two directions were 13.58\%, and that for confused voxels was 0.64\%. The head model was finally computed using Eq.~(\ref{eq03}). 

%======================
% Figure-7
%======================

\begin{figure*}
\includegraphics[width=\textwidth]{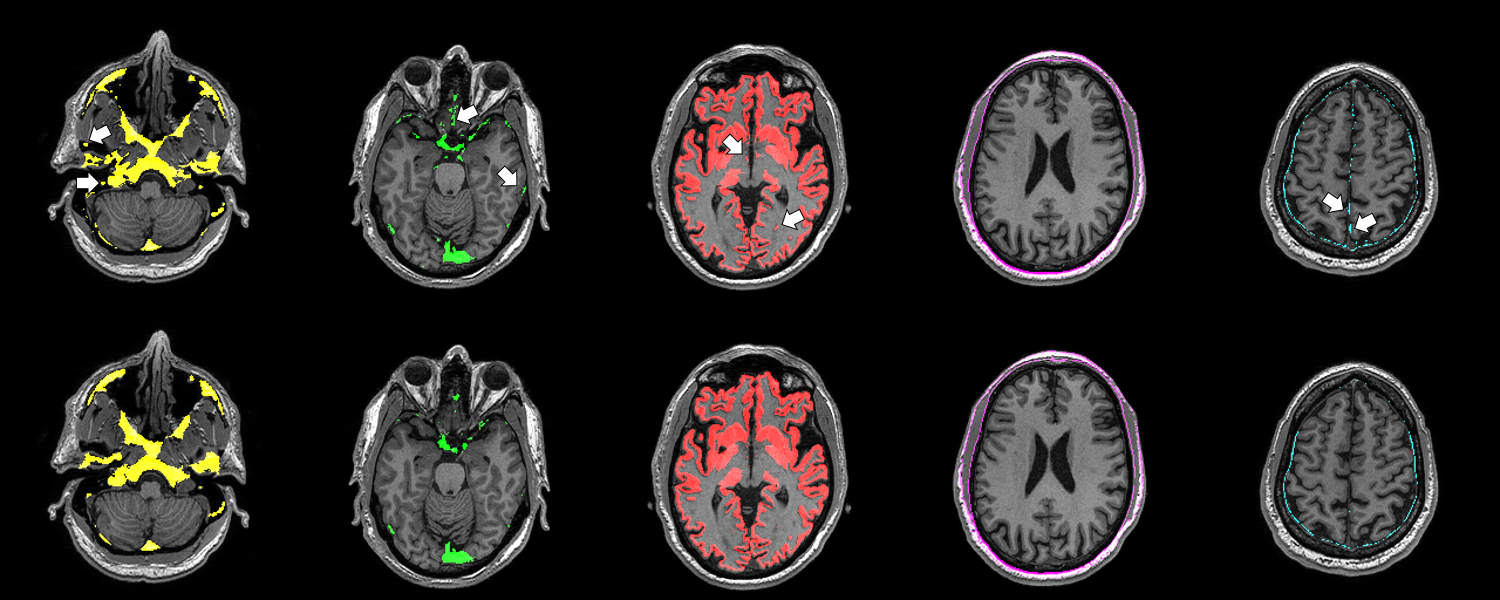}
\caption{Comparison demonstration of the original head model $R^\circ$ (top) and the network-generated head model $R^\psi$ (bottom) overlapped over their corresponding MRI slices. From left to right, color labels identify bone (canc.), blood vessels, GM, fat, and dura. Arrows in the top row indicate some regions where labels are mismatched.}
\label{P05_overlap}
\end{figure*}

Dice coefficients computed for different anatomical structures are shown in Table~\ref{TabDice}. From this table, it can be seen that the highest segmentation values were observed for the WM and GM, owing to their high contrast and being well-defined tissues in MRI scans. However, low scores were obtained for dura and blood vessels. This tendency was expected owing to their limited spatial distribution as well as their low contrast with surrounding tissues. To analyze the network-generated model and compare it with the original one, selected tissues are shown overlapped on the MRI image shown in Fig.~\ref{P05_overlap}.

%=================
% TABLE-3
%=================

\begin{table*}
\centering
\footnotesize
\caption{Mean and standard deviation of Dice coefficient values computed for 18 subjects of different network-generated models. Bold indicate superior mean value.}
\label{TabDice}
\setlength{\tabcolsep}{3pt}
\begin{tabular}{| l| c  c  c c c|}
\hline
{\bf Tissue}	& U-net & $R^\alpha$ & $R^\beta$ & $R^\gamma$ & $R^\psi$  \\
\hline
\hline
Skin   &85.23$\pm$4.36 &90.21$\pm$5.36 	& 90.19$\pm$6.08	& 90.25$\pm$5.12	&{\bf 91.51}$\pm$5.35\\
Muscle&83.19$\pm$6.84 & 91.21$\pm$6.16&	91.96$\pm$5.03&	91.93$\pm$5.76	&{\bf 92.90}$\pm$5.24\\
Fat   &80.58$\pm$7.80 & 80.85$\pm$8.42	&81.33$\pm$8.57&	80.64$\pm$7.01	&{\bf 83.18}$\pm$7.72\\
Bone (Cort.)& 84.10$\pm$5.39&85.68$\pm$5.44	&85.55$\pm$5.70	&85.69$\pm$6.18&	{\bf 87.57}$\pm$5.52\\
Bone (Canc.)&76.95$\pm$9.89 & 81.24$\pm$9.03	&82.13$\pm$8.66&	80.15$\pm$12.85	&{\bf 84.18}$\pm$8.66\\
Dura & 45.62$\pm$10.84&{\bf 56.77}$\pm$8.81 &	47.82$\pm$11.79&	46.36$\pm$12.50	&55.13$\pm$10.17\\
Blood &73.29$\pm$9.03& 73.11$\pm$10.94&	69.24$\pm$10.15	&72.28$\pm$12.03	&{\bf 75.20}$\pm$10.08\\
CSF &86.82$\pm$4.23& {\bf 90.38}$\pm$4.55	&82.86$\pm$6.36&	82.45$\pm$5.77	&88.30$\pm$4.82\\
GM & 89.67$\pm$3.34&{\bf 95.79}$\pm$3.31	&88.96$\pm$3.15	&87.42$\pm$5.04&	93.61$\pm$2.75
\\
WM &92.76$\pm$4.74& {\bf 96.74}$\pm$2.28&	93.65$\pm$2.34	&93.15$\pm$1.59	&95.96$\pm$2.05\\
Cerebellum & 91.85$\pm$2.00&93.72$\pm$2.43	&92.00$\pm$2.40	&91.78$\pm$2.07	&{\bf 93.85}$\pm$2.05\\
V. Humor &93.75$\pm$2.64&95.93$\pm$2.60	&95.99$\pm$2.94	&96.11$\pm$2.45&	{\bf 96.95}$\pm$2.39 \\
Mucous tissue&67.50$\pm$16.25&80.02$\pm$13.78	&80.08$\pm$13.98&	80.59$\pm$12.78	&{\bf 82.31}$\pm$13.06\\
\hline
\end{tabular}
\end{table*}

%======================
% Figure-8
%======================

%======================
% Figure-9
%======================

\begin{figure*}
\includegraphics[width=\textwidth]{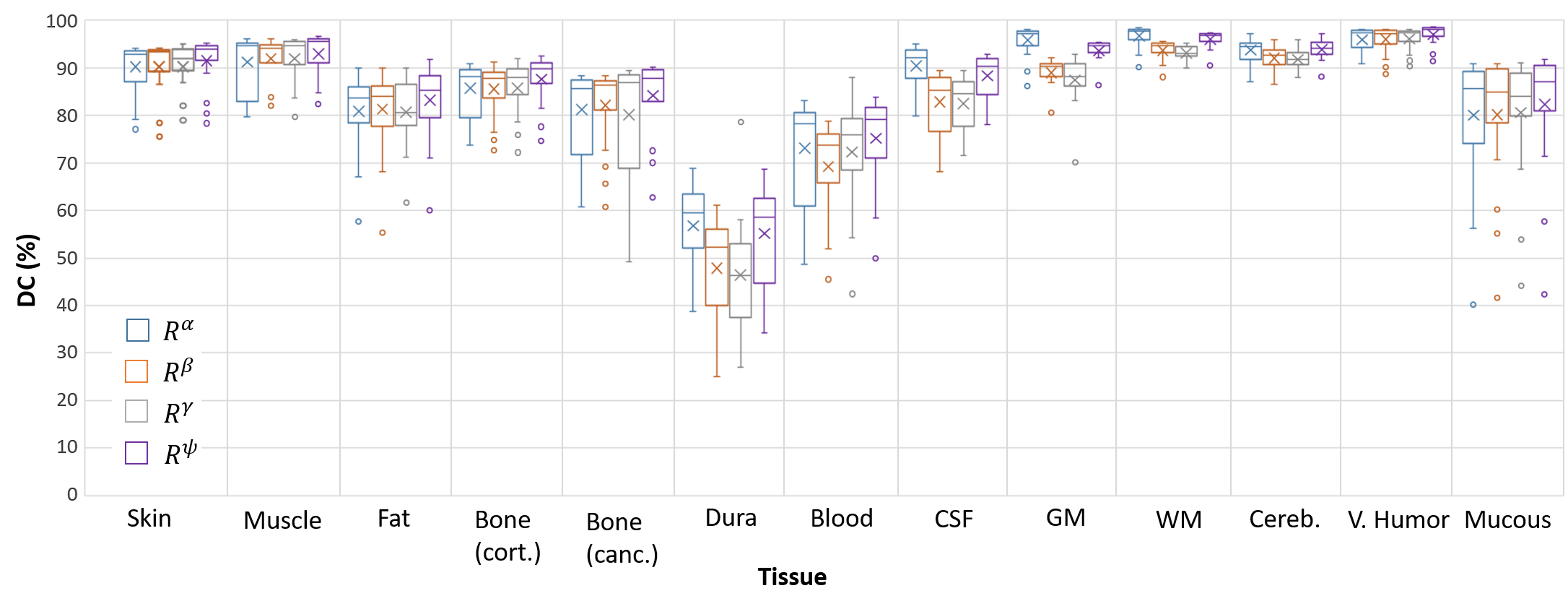}
\caption{Boxplot of Dice coefficients computed from 18 subjects using ForkNet for different head structures. Cross indicate the mean value, center line indicate the median, box indicate first and third quartiles, whiskers indicate maximum and minimum values and circles indicate outliers. Outliers are defined as the values that falls more than 1.5 times the interquartile range above (or below) the third (first) quartile. A sample of original and network-generated head models are shown in Supplementary Figures \ref{case01015}-\ref{case01045}.}
\label{Dice_all}
\end{figure*}

The above-mentioned experiment was repeated using different subjects to validate the robustness of the network to subject variability. The Dice coefficients for all segmented tissues are shown in Fig.~\ref{Dice_all}. These results demonstrate the effects of subject variability on the quality of ForkNet segmentation. Sample of the corresponding head models are shown in Supplementary Figs. \ref{case01015}-\ref{case01045}. On the one hand, the Dice coefficients of the WM, GM, cerebellum, and vitreous humor reached high values (the average was higher than 90\%), which can be expected owing to the high-contrast anatomical representation of these tissues in MRI. On the other hand, the dura (average of less than 50\%) and mucous tissue (average of less than 60\%) were the tissues with the lowest accuracy. This can be attributed to the fact that dura is usually present as thin layer (small number of voxels) and mucous tissue commonly appears with low contrast in MRI images. By looking at the results shown in Fig.~\ref{Dice_all}, it is observed that  single outliers presented in WM, GM, Fat, and Muscle are corresponding to a single subject. This may indicate a strong bias in MRI data corresponding to that subject. Other outliers are mainly related to segmentation error in relatively small size structures. As shown below in Section 4.3, we have found that the effect of segmentation error of dura and mucous is insignificant to electromagnetic stimulation. In previous studies (e.g., \cite{Huang2013JNE}), those tissues are not included in human head modeling.

%======================
% Figure-10
%======================

\begin{figure*}
\centering
\includegraphics[width=\textwidth]{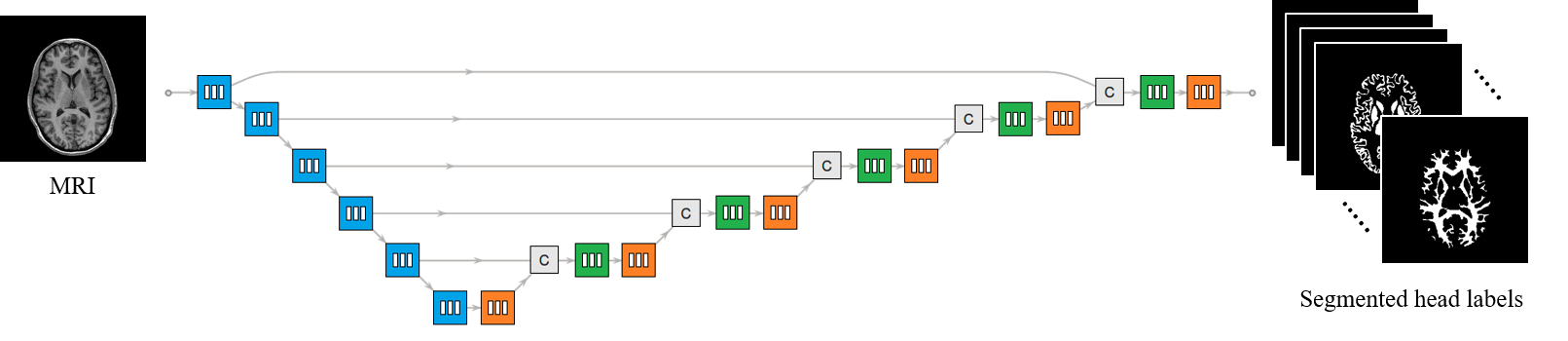}
\caption{Architecture of the corresponding U-Net model used for obtaining the results shown in Fig.~\ref{unet}. The detailed computation of the feature variables for each layer is presented in Table~\ref{UnetSize}.}
\label{UNetArch}
\end{figure*}

%======================
% Table-4
%======================

\begin{table*}
\centering
\footnotesize
\caption{Details of the U-Net architecture shown in Fig.~\ref{UNetArch}.}
\begin{tabular}{|l|lllll|}
\hline 
{\bf Module} & {\bf Layer} & {\bf Output size} & {\bf Kernel} & {\bf Stride} & {\bf Padding}\\
\hline \hline
Input & & $2^{8}\times2^{8}$ &&&\\
\hline
EncMod$_{i}$ & Convolution & $2^{(i+2)} \times 2^{(9-i)} \times 2^{(9-i)}$ & ($2^{(i+2)}$ $\times$3$\times$3) & (1$\times$1) & (1$\times$1)\\
$i=1 \rightarrow 6$& BN \& ReLU & $2^{(i+2)} \times 2^{(9-i)} \times 2^{(9-i)}$ & &  & \\
& Pooling (Max) & $2^{(i+2)} \times 2^{(8-i)} \times 2^{(8-i)}$ &&& \\
\hline
DecMod$_{j}$ & Deconvolution & $2^{(j+1)} \times 2^{(9-j)} \times 2^{(9-j)}$ & ($2^{(j+1)}$ $\times$2$\times$2) & (2$\times$2) & \\
$j=6 \rightarrow 2$ & BN \& ReLU & $2^{(j+1)} \times 2^{(9-j)} \times 2^{(9-j)}$ & &  & \\
 & Convolution & $2^{(j+1)} \times 2^{(9-j)} \times 2^{(9-j)}$ & ($2^{(j+1)}$ $\times$3$\times$3) & (1$\times$1) & (1$\times$1)\\
\hdashline
DecMod$_{j}$ & Deconvolution & $13 \times 2^{(9-j)} \times 2^{(9-j)}$ & (13$\times$2$\times$2) & (2$\times$2) & \\
$j= 1$ & BN \& ReLU & $13 \times 2^{(9-j)} \times 2^{(9-j)}$ & &  & \\
 & Convolution & $13 \times 2^{(9-j)} \times 2^{(9-j)}$ & (13$\times$3$\times$3) & (1$\times$1) & (1$\times$1)\\
\hline
ConvMod$_{j}$ & Convolution & $2^{(j+2)} \times 2^{(8-j)} \times 2^{(8-j)}$ & ($2^{(j+2)}$ $\times$3$\times$3) & (1$\times$1) & (1$\times$1)\\
$j=5 \rightarrow 2$ & BN \& ReLU & $2^{(j+2)} \times 2^{(8-j)} \times 2^{(8-j)}$ & &  & \\
\hdashline
ConvMod$_{j}$ & Convolution & $13 \times 2^{(8-j)} \times 2^{(8-j)}$ & (13$\times$3$\times$3) & (1$\times$1) & (1$\times$1)\\
$j= 1$ & BN \& ReLU & $13 \times 2^{(8-j)} \times 2^{(8-j)}$ & &  & \\
\hline
Concat$_{j}$ & Concatenation &  $2^{(j+3)} \times 2^{(8-j)} \times 2^{(8-j)}$ & &&\\
$j=5 \rightarrow 1$ &  &  & &&\\
\hline
Output & & $ 13 \times 2^{8}\times2^{8}$ &&&\\
\hline
\end{tabular}
\label{UnetSize}
\end{table*}

%==========================
% 4.2 Comparison with U-net architecture
%==========================

\subsection{Comparison with the U-net architecture}

\begin{figure*}
\center
\includegraphics[width=\textwidth]{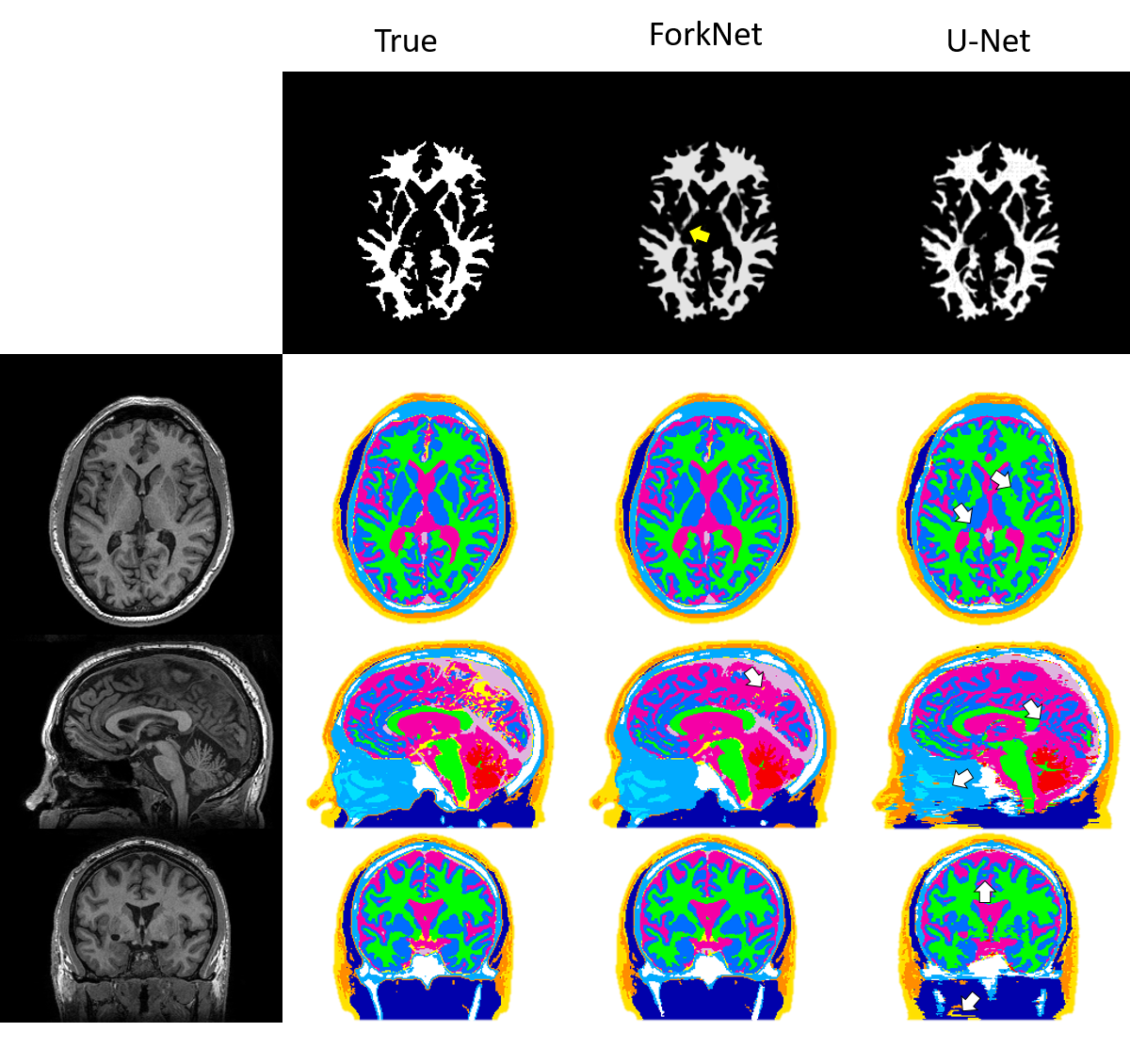}
\caption{Segmentation results using the ForkNet and U-net architectures for a single tissue (WM) in the top row and for all head tissues in the bottom row. Arrows indicate some regions where labels are incorrect.}
\label{unet}
\end{figure*}

To understand how the proposed architecture performs differently from the U-net architecture \citep{Ronneberger2015MICCAI}, we conducted an experiment using the corresponding U-net architecture in two segmentation cases. In the first case, we consider the segmentation of a single anatomical structure (WM), and labels were prepared by marking WM as foreground and all the other tissues as background. The second case considered the segmentation of all 13 head tissues. We prepared the corresponding U-net architecture with its all parameters identical to those of ForkNet in both cases. The U-net architecture used in this study is shown in Fig.~\ref{UNetArch} and detailed features are in Table~\ref{UnetSize}. The cross-entropy loss function minimized by ADAM algorithm is used for training with 50 iterations and batch size=2. We have used exactly the same parameters for both networks for unbiased evaluation. Both networks were trained using the axial direction data, and the segmentation results are shown in Fig.~\ref{unet}. In the first case, in which the segmentation problem is quite simple, we can appreciate the superior quality offered by the U-net architecture. However, when considering more anatomical structures in the second case, the performance of the U-net architecture was overshadowed by that of the ForkNet architecture. Several artifacts and incorrect labeling can be observed in the U-net results.

\begin{figure*}
\includegraphics[width=\textwidth]{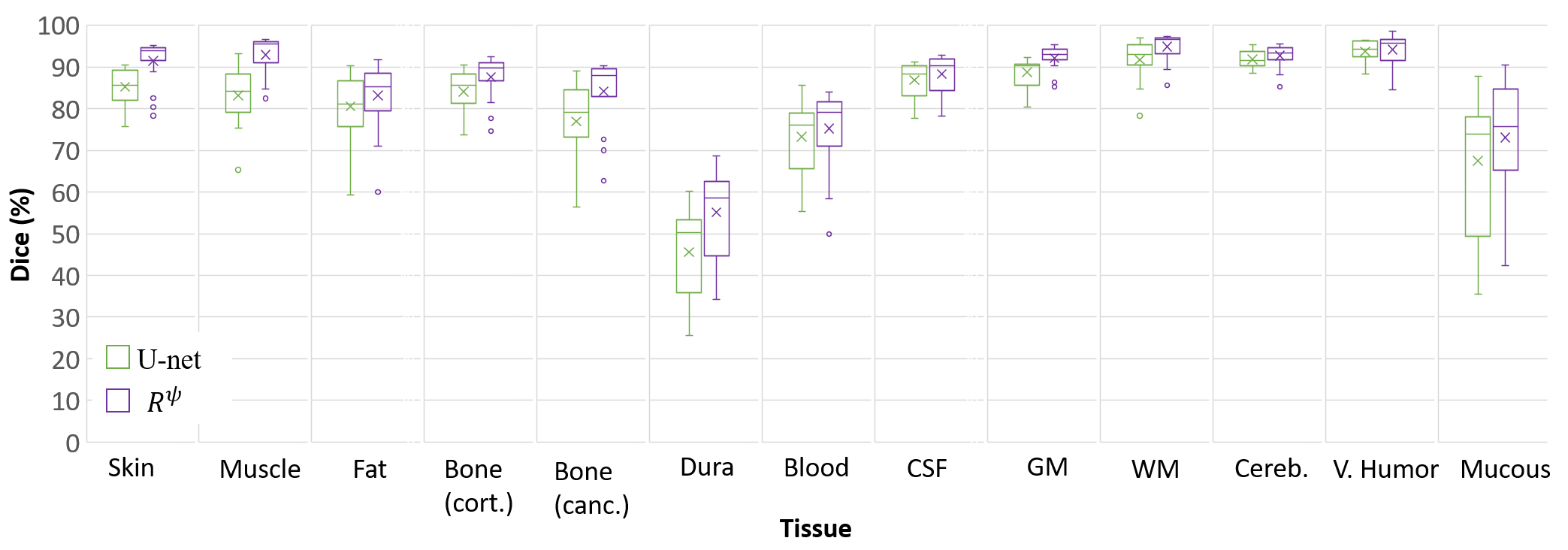}
\caption{Boxplot of Dice coefficients computed from 18 subjects using U-net and ForkNet for different head structures.}
\label{Dice_all_Unet}
\end{figure*}

\begin{figure*}
\includegraphics[width=\textwidth]{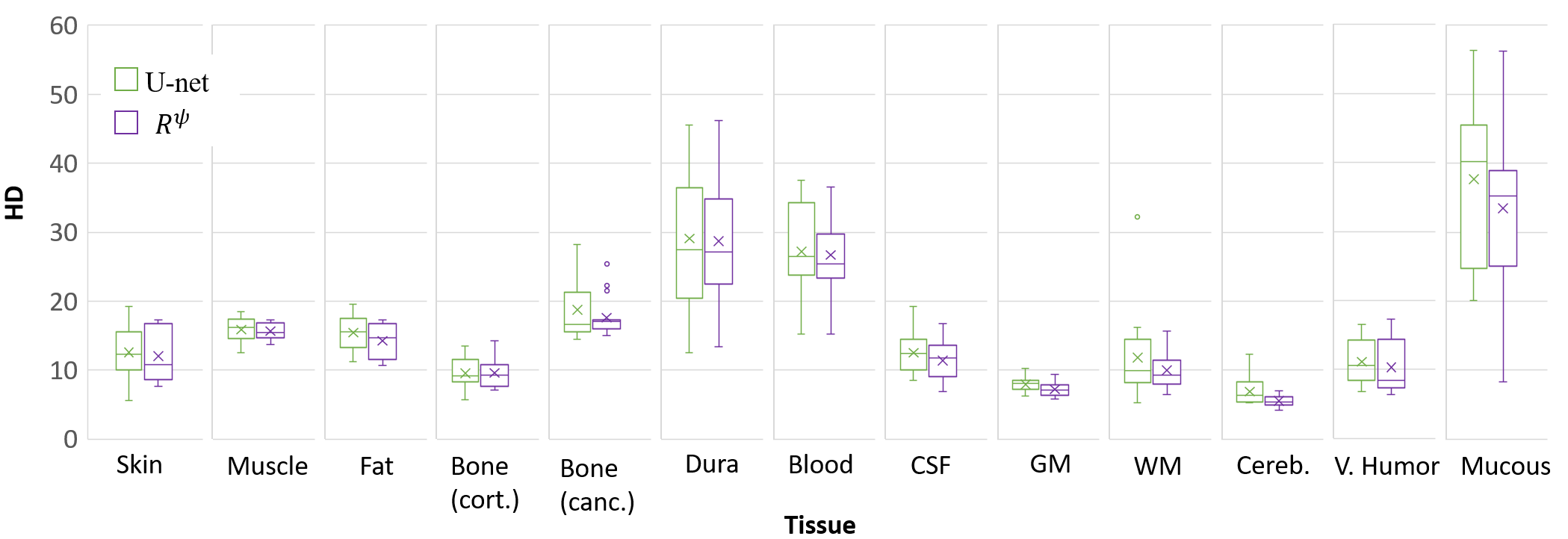}
\caption{Boxplot of Hausdorff distance computed from 18 subjects using U-net and ForkNet for different head structures.}
\label{HD_all}
\end{figure*}

This experiment is repeated with all the 18 subjects using leave-one-out cross-validation and the DC is computed for both architectures and results are shown in Fig.~\ref{Dice_all_Unet} and HD values are shown in Fig.~\ref{HD_all}. HD is a segmentation quality measure that is more sensitive to voxel position compared to DC. It highlighted the segmentation faults efficiently especially for large size structures. A high HD value may exist if, for example, a single incorrect voxel is segmented relatively far from the original position. However, HD value may not be consistent with the TMS EF distribution, which is hardly being effected by segmentation fault of a single (or few voxels). HD measurements shown in Fig.~\ref{HD_all} demonstrate good segmentation results, especially for brain tissues. Again, dura, blood vessels, and mucous tissues are presented in relatively high HD values. These results indicate that the ForkNet architecture outperforms the conventional U-net architecture in the segmentation results of almost all structures.

%======================
% Figure-11
%======================

%======================
% Figure-12
%======================

%======================
% Figure-13
%======================

%==========================
% 4.3 TMS simulation
%==========================

\subsection{TMS simulation}

\begin{figure*}
\centering
\includegraphics[width=\textwidth]{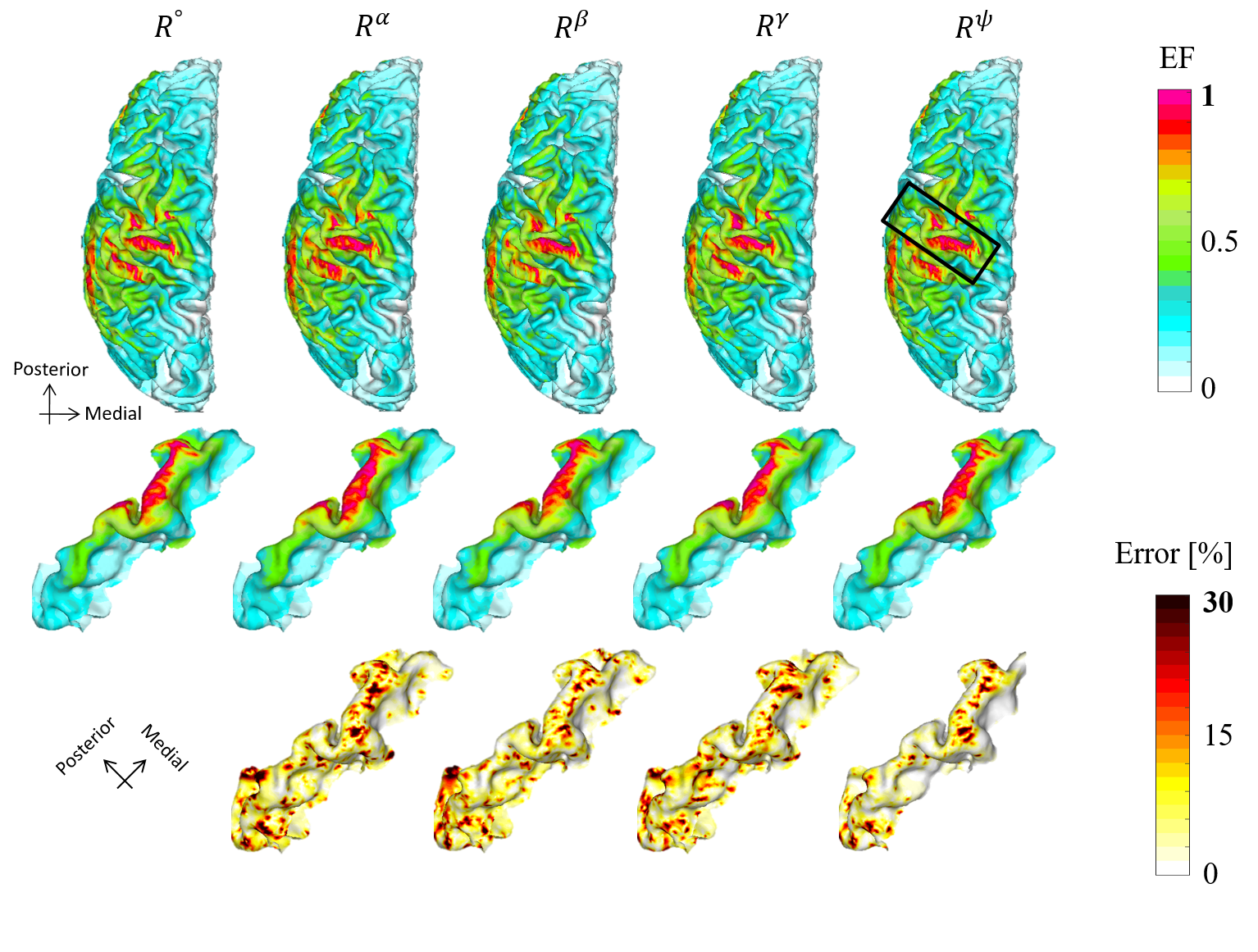}
\caption{Normalized electric field distribution map in the brain (axial view) using the original head model $R^\circ$ and the network-generated models $R^{\alpha}$, $R^{\beta}$, $R^{\gamma}$, and $R^{\psi}$ for subject (case01020). The region labeled with a black rectangle is magnified and shown below along with error maps.}
\label{tms5}
\end{figure*}

The five head models, namely the original $R^\circ$ and the network-generated models $R^{\alpha}$, $R^{\beta}$, $R^{\gamma}$, and $R^{\psi}$, were used to carry out TMS stimulations. The electric fields induced via TMS in the brain cortex is shown in Fig.~\ref{tms5}, computed using the generated head models and the ground truth. For comparison, a percentage error map is shown in the target area (hand motor area). Error is computed considering the head model generated from the semi-automatic method ($R^\circ$) as the golden truth. The results for five subjects are summarized in Table~\ref{TabTMS1} in terms of the MAE metric. These results indicate that the simulation results obtained using the network-generated models (especially $R^{\psi}$) were highly consistent with the original segmented model. Although some segmentation errors can be observed, especially in the dura and mucus (Fig.~\ref{Dice_all}), their impact in the TMS simulation results was insignificant. The TMS simulation is repeated to another four subjects to investigate subject variability in TMS stimulation when network generated models are used.

\begin{figure*}
\centering
\includegraphics[width=\textwidth]{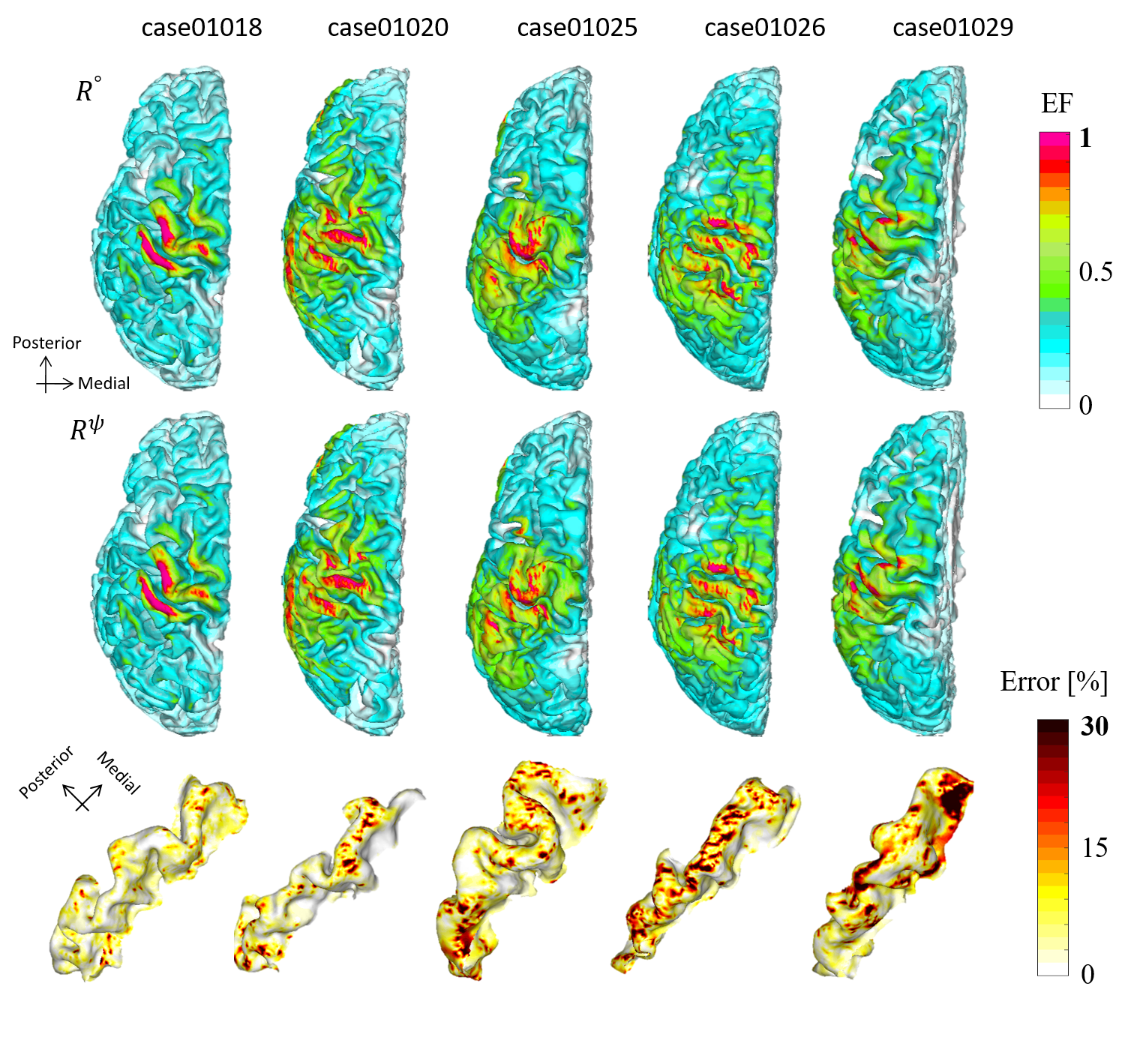}
\caption{Normalized electric field distribution map in the brain using the original head model $R^\circ$ and the network-generated model $R^{\psi}$ for five subjects. The corresponding error maps of the hand motor cortex are shown below.}
\label{tms6}
\end{figure*}

The results shown in Fig.~\ref{tms6} indicate a relatively small subject variability in TMS simulations, and the resulting point-to-point electric field distribution errors are listed in Table~\ref{TabTMS1}. These results demonstrate that a small error could be attained in both the target area and the hotspot within the target area. In few regions, a patchy-like artifacts with error around 30\% can be observed. These can be mainly due to the local difference in the segmentation of the GM/CSF. TMS is known to be very sensitive to high contrast difference in the surface. Moreover, the lack of dura closedness might have a small potential impact on the electrictromagnetic exposure.

%======================
% Figure-14
%======================

%======================
% Figure-15
%======================

%=================
% TABLE-5
%=================

\begin{table}
\centering
\footnotesize
\caption{Error metrics in the motor area (MAE) and the hotspot in the motor area (MAE$_{0.7}$) of the TMS simulation data shown in Figs.~\ref{tms5} and \ref{tms6}.}
\label{TabTMS1}
\setlength{\tabcolsep}{3pt}
\begin{tabular}{| l| c| c | c| }
\hline
{\bf Subject}& {\bf Model}	&  MAE& MAE$_{0.7}$ [\%] \\
\hline
\hline
\multirow{4}{*}{case01020} &$R^\alpha$ & 2.2 & 7.4\\
\cline{2-4}
& $R^\beta$ & 1.9 & 5.6\\
\cline{2-4}
& $R^\gamma$ & 2.2 & 6.9\\
\cline{2-4}
& $R^\psi$ & 1.3 & 5.6\\
\hline
case01018 & $R^\psi$ & 1.2 & 4.0\\
\hline
case01025 & $R^\psi$ & 2.6 & 5.3\\
\hline
case01026 & $R^\psi$ & 2.7 & 7.3\\
\hline
case01029 & $R^\psi$ & 3.2 & 6.2\\
\hline
\end{tabular}
\end{table}

%==========================
% 5. Discussion
%==========================

\section{Discussion}

%==========================
% 5.1 ForkNet architecture
%==========================

\subsection{ForkNet architecture}

A common segmentation challenge is the identification of the background when it is a composite of heterogeneous anatomy. This problem is discussed in the work of \cite{Wachinger2018NeuroImage}, where a solution is provided using hierarchical segmentation (DeepNAT). In this method, two-step segmentation is performed using two networks; the first network is used to identify the background, followed by anatomical segmentation. The results on the performance of DeepNAT are interesting in terms of ability to remove background using a separate network. However, both networks should be trained carefully because segmentation errors in the first network may accumulate in the second. The ForkNet architecture proposed in this paper can handle this problem efficiently by employing independent decoders for each anatomical structure. Within a single decoder, the corresponding anatomy is considered as foreground and all the other structures are considered as background. Feed-forward from the encoders at different resolution levels enables the optimization of features with high accuracy. We believe that this is a key reason for the observed performance of ForkNet.

%======================
% Figure-16
%======================

\begin{figure*}
\centering
\includegraphics[width=\textwidth]{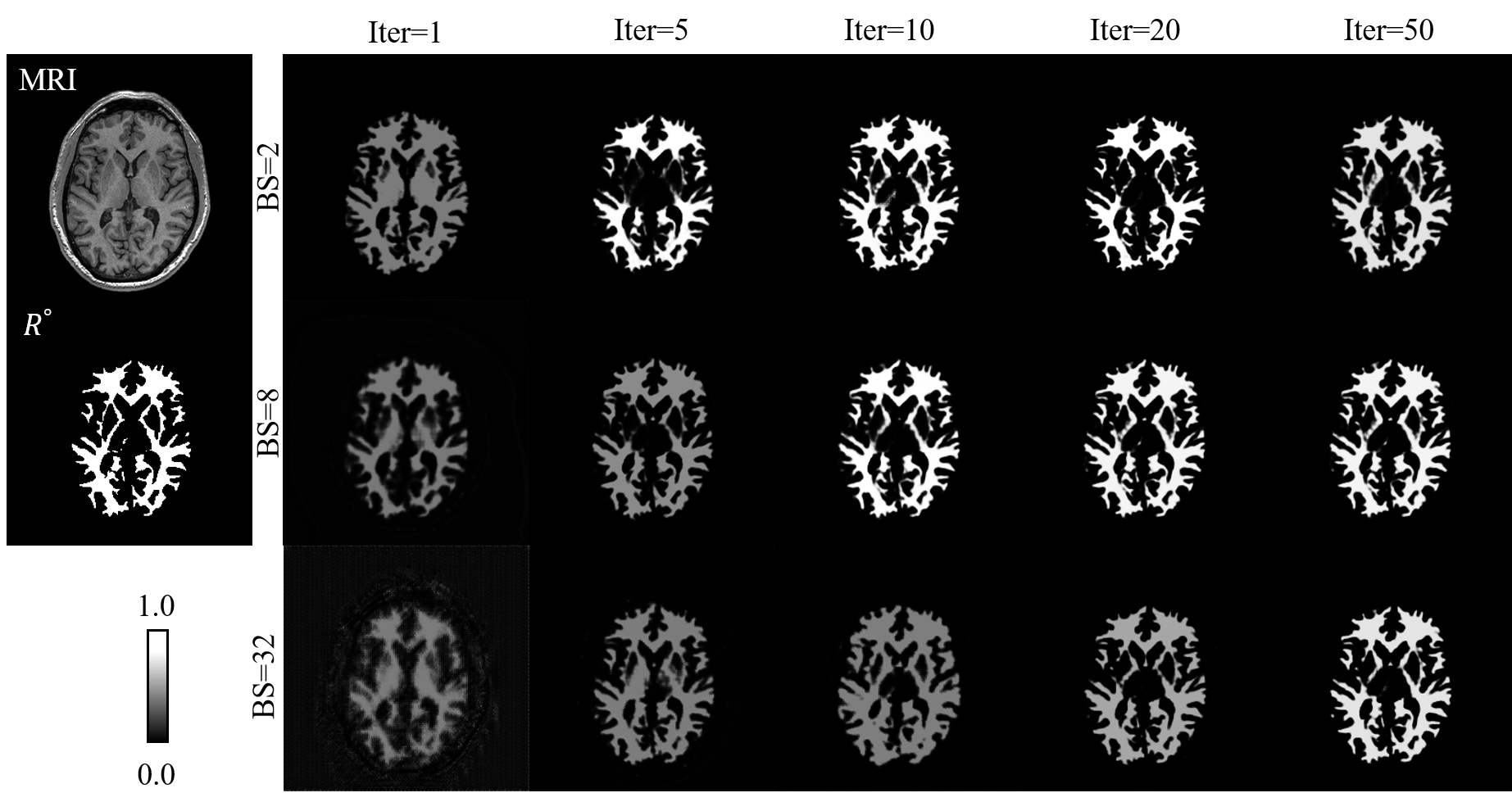}
\caption{Central axial MRI image ($M_k$) and the corresponding original segmentation labels for white matter ($R^\circ_{k,n}$). The network output ($L_{k,n}$) was computed with training phases consisting of 1, 5, 10, 20, and 50 iterations and batch sizes (BSs) of 2, 8, and 32. Labeled images are shown in gray scale [0.0, 1.0].}
\label{analysis}
\end{figure*}

\begin{figure*}
\center
\includegraphics[width=.8\textwidth]{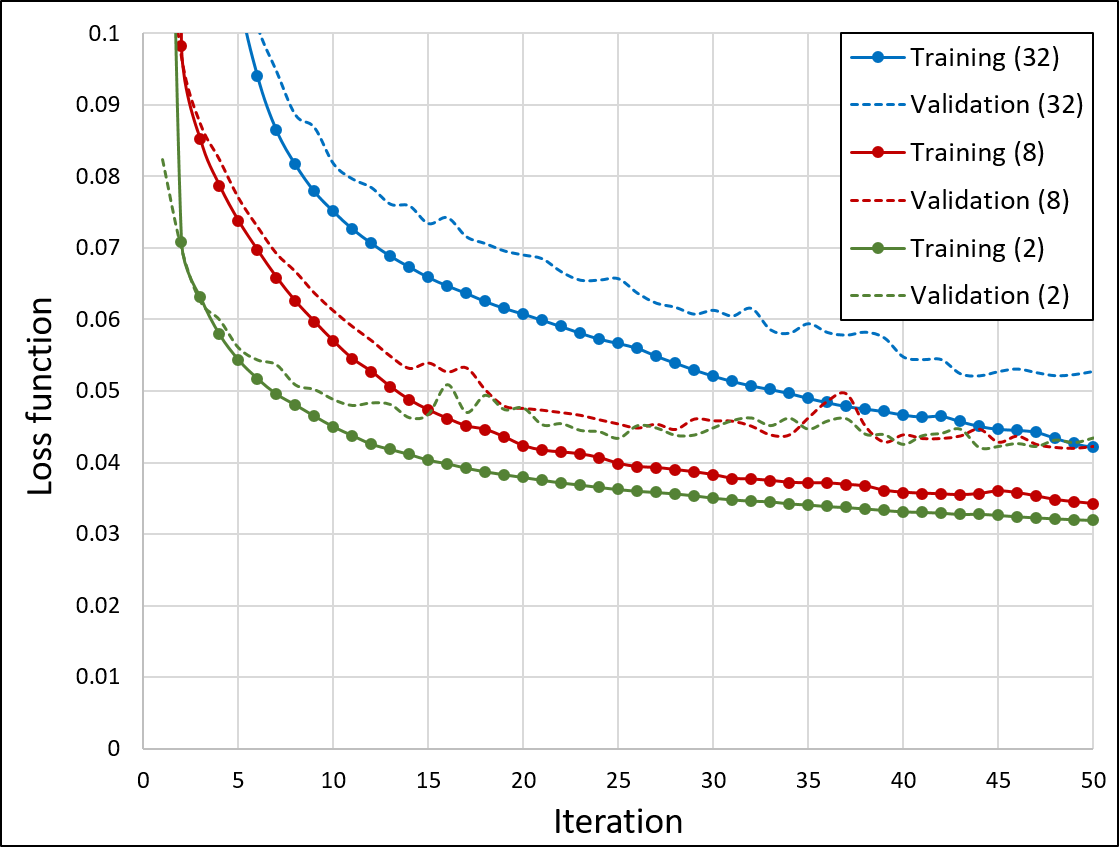}
\caption{Loss function computed for training and validation using different batch sizes in the axial direction for the white matter structure shown in Fig.~\ref{analysis}.}
\label{iter_sb}
\end{figure*}

One major obstacle in deep learning applications is the optimization of network parameters for specific architectures used for naming applications. With the recent interest in semantic segmentation using deep learning, several appealing architectures have been proposed. Although there are some general architectures that can work efficiently in almost every application, the performance of some of these architectures under specific conditions are still unclear. Because the architecture presented in this study is novel, it would be helpful to demonstrate the performance of the network for parameter variations. For simplicity, we traced the network’s performance for segmenting a single anatomical structure (WM) using the axial slicing direction, and an experiment was conducted with different numbers of training iterations and batch sizes. The segmentation results for the central axial slice are shown in Fig.~\ref{analysis}, and a plot of the loss function for training and validation is shown in Fig.~\ref{iter_sb}. It is clear that segmentation quality convergence is inversely proportional to batch size. Segmentation certainty (the ability to derive crisp labels) is higher for a smaller batch sizes. However, network accuracy (correct labeling) behaves in the opposite way. We found that small improvement can be achieved after approximately 50 iterations. These results are consistent with the loss function measurements shown in Fig.~\ref{P05_Loss}.

%======================
% Figure-17
%======================

%======================
% 5.2 Training data limitations
%======================

\subsection{Training data limitations}

Within the framework of supervised-based segmentation, the performance of the developed approaches is highly related to the quality of the training dataset. Several parameters such as the size of the available dataset, accuracy of the segmentation, variety of subjects and imaging modality, etc. are known to have a high impact on the training guidance towards a realistic unbiased automatic segmentation. Several excellent high-quality datasets are available for brain segmentation (some are listed here\footnote{https://grand-challenge.org/challenges/}). However, to the best of our knowledge, there is no freely available dataset for all head segmentation.

The semi-automatic method \citep{Laakso2015BS}, that is used here to generate the training data, is the state of the art for full head segmentation. It is proved a high matching results between the estimated threshold from computed electric field using the semi-automatic method and real measurements in TMS and tDCS \citep{Laakso2018BS, Mikkonen2018FN,Aonuma2018neuroimage}. As the semi-automatic method has some tuning parameters that need to be adjusted, it is expected that it may provide some segmentation faults, especially for non-brain regions, that appears in low-contrast in MRI. This limitation is expected as accurate segmentation of non-brain tissues (for example bones and blood vessels) requires additional anatomical information that can be provided using CT and/or venogram. Segmentation using semi-automatic method of different subjects are shown in the Supplementary Figs. \ref{case01015}-\ref{case01045} for assessment. Once a higher quality head segmentation dataset is available, it is possible for ForkNet to be re-trained to improve the automatic segmentation accuracy.

%==========================
% 5.3 Robustness validation
%==========================

\subsection{Robustness validation}
The main motivation of this study is to provide a robust personalized head modeling that has a potential clinical usefulness. One major challenge within this prospective is the large varieties of scanning parameters such as scanner magnet power, resolution, and other physical modeling parameters. It is well-known that MRI is a highly biased imaging modality with high-intensity variability even with unified subject and scanner. The development of automatic segmentation method that can be used in different scanners is challenging and requires several pre-processing steps to handle bias effect and other scanner physical characteristics. In such cases, the proposed ForkNet can be re-trained using data obtained from specific scanner for clinical use. It worth noting that obtaining the golden truth data required for re-training is known to be a time consuming process especially if the annotation/segmentation is performed manually. Another option is the use of transfer learning to adopt the learned features to fit with new scanner data (e.g. \cite{VanOpbroek2019TMI}).  

%======================
% Figure-18
%======================

\begin{figure*}
\center
\includegraphics[width=\textwidth]{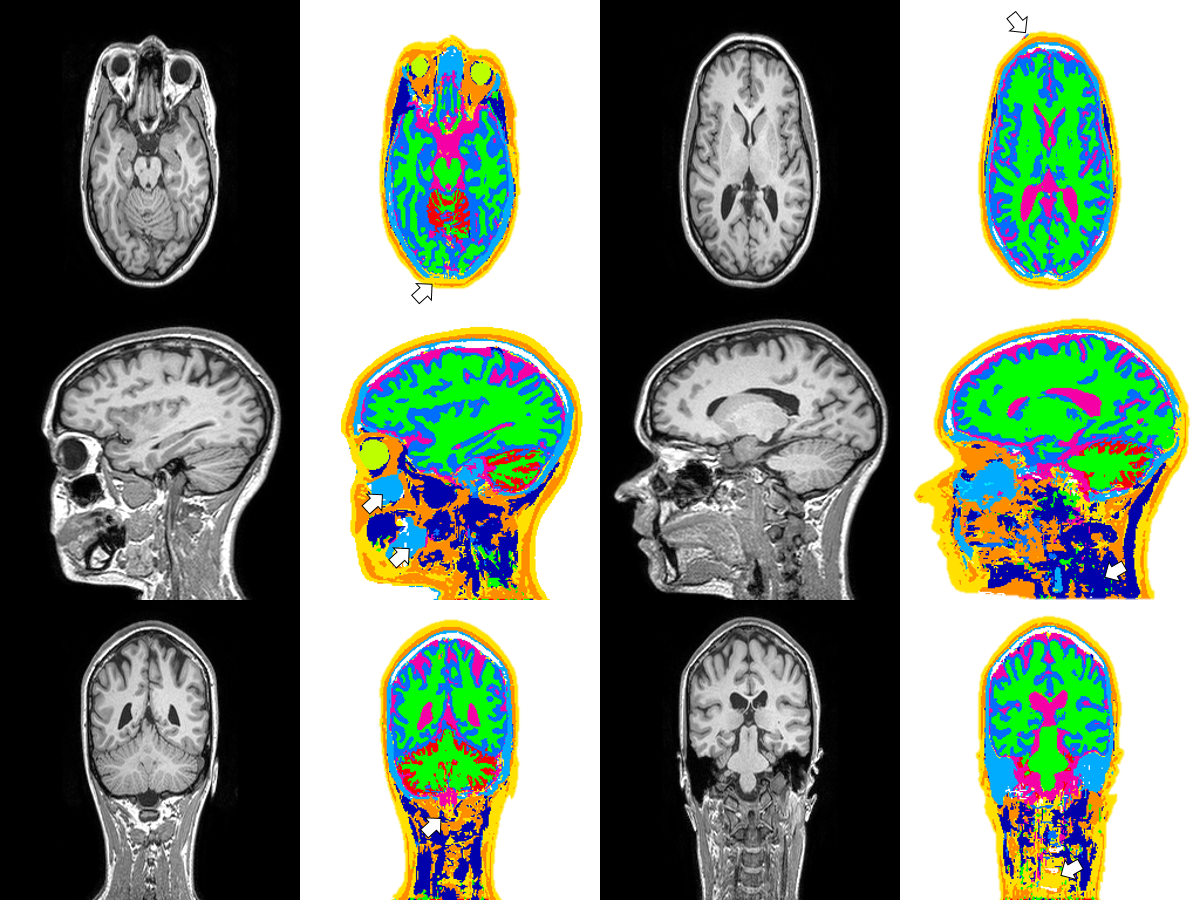}
\caption{MRI and corresponding ForkNet segmented head models generated for subject IXI025 from IXI (Guys) dataset. Axial, sagittal, and coronal slices are arranged from top to bottom and arrows indicate some regions where segmentation is inaccurate.}
\label{IXIGuys025}
\end{figure*}

%======================
% Figure-19
%======================

\begin{figure*}
\center
\includegraphics[width=\textwidth]{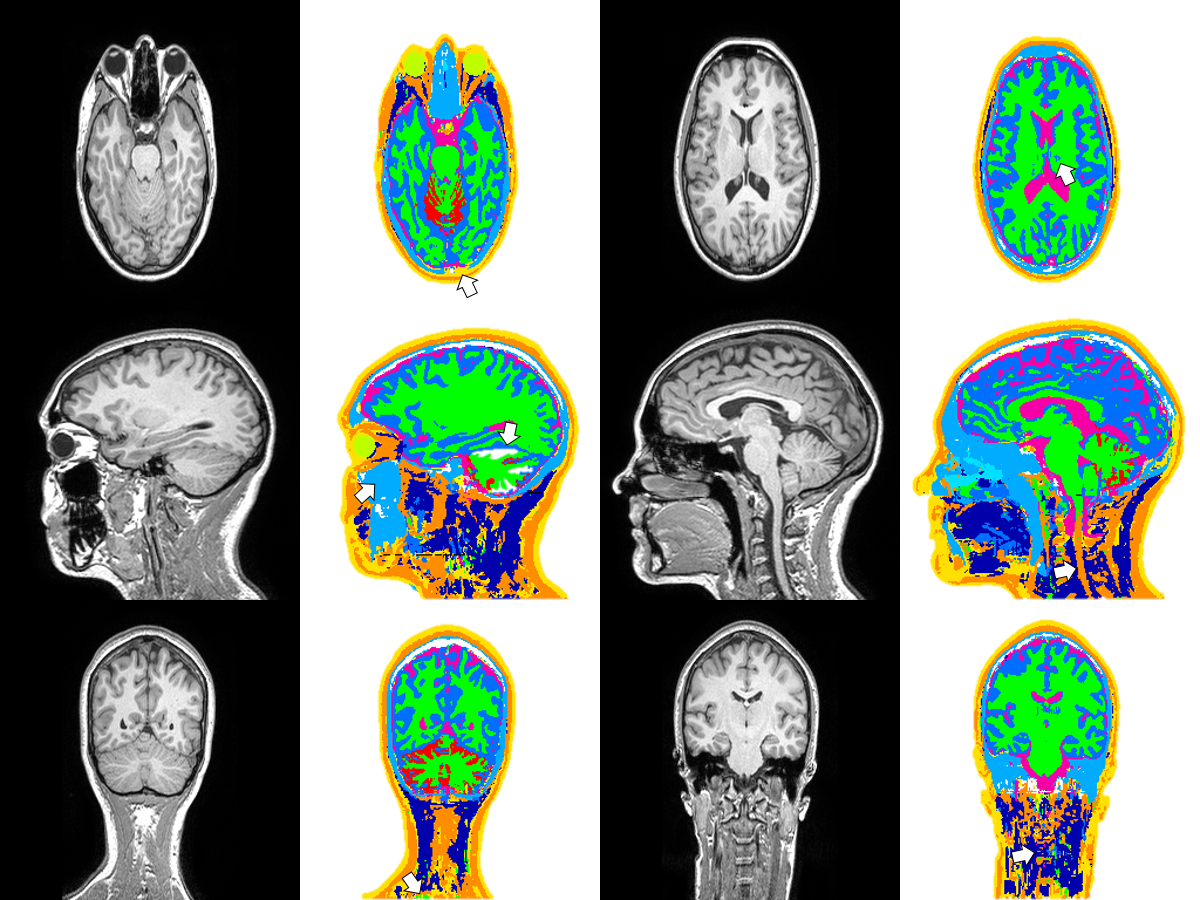}
\caption{MRI and corresponding ForkNet segmented head models generated for subject IXI068 from IXI (Guys) dataset. Axial, sagittal, and coronal slices are arranged from top to bottom and arrows indicate some regions where segmentation is inaccurate.}
\label{IXIGuys068}
\end{figure*}

\begin{figure*}
\center
\includegraphics[width=\textwidth]{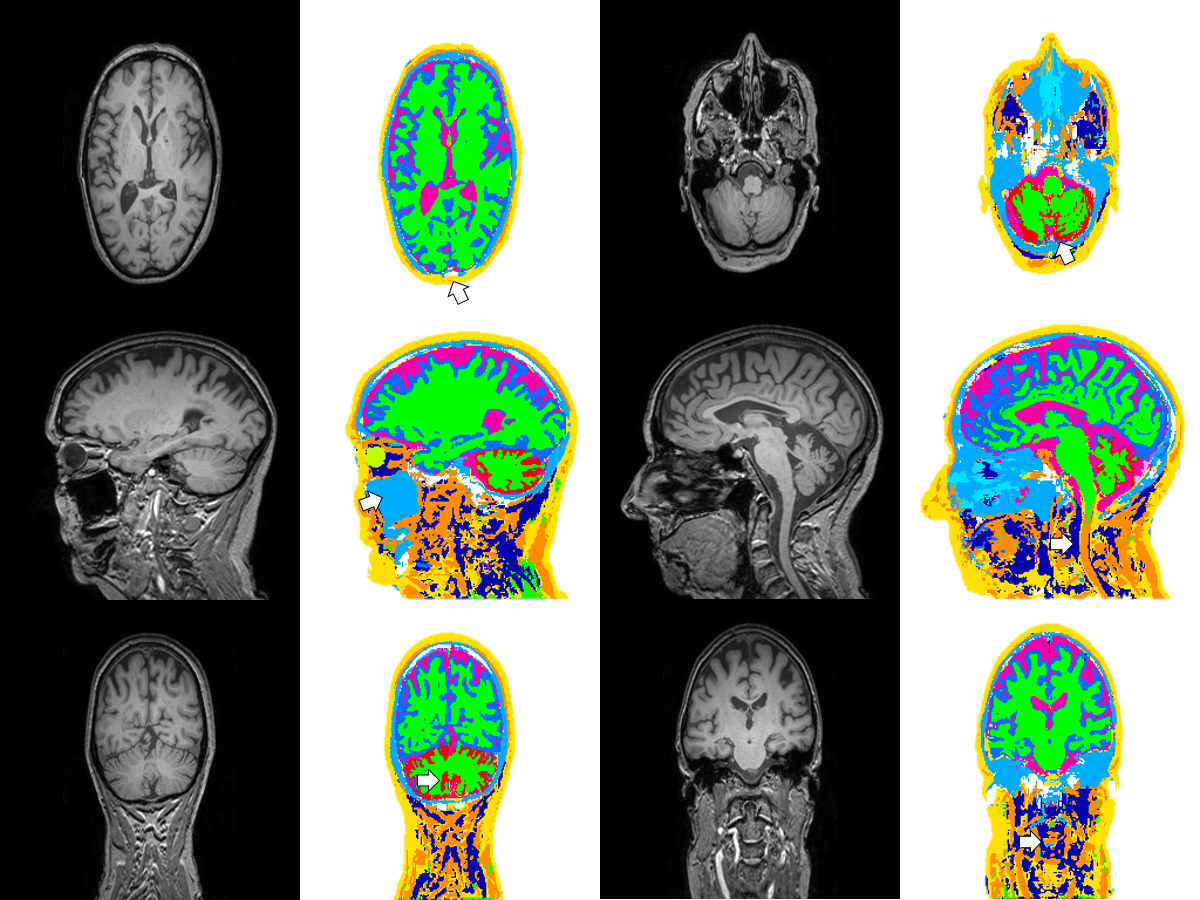}
\caption{MRI and corresponding ForkNet segmented head models generated for subject IXI518 from IXI (HH) dataset. Axial, sagittal, and coronal slices are arranged from top to bottom and arrows indicate some regions where segmentation is inaccurate.}
\label{IXIHH518}
\end{figure*}

\begin{figure*}
\center
\includegraphics[width=\textwidth]{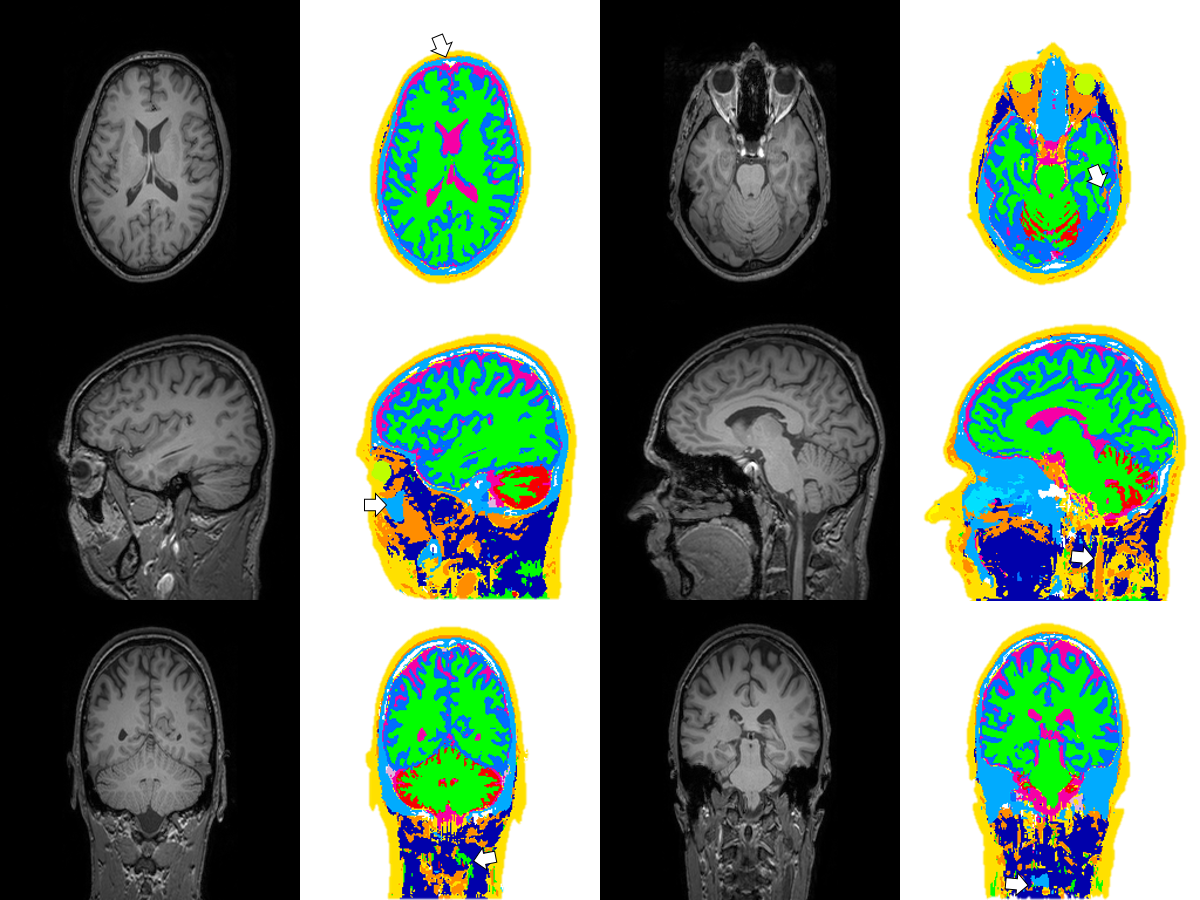}
\caption{MRI and corresponding ForkNet segmented head models generated for subject IXI519 from IXI (HH) dataset. Axial, sagittal, and coronal slices are arranged from top to bottom and arrows indicate some regions where segmentation is inaccurate.}
\label{IXIHH519}
\end{figure*}

However, we conduct another study to validate the robustness of ForkNet using data obtained from different MRI scanners. The IXI Dataset\footnote{http://brain-development.org/ixi-dataset/} contains a collection of nearly 600 MRI from normal healthy subjects. We evaluated the use of trained networks using NAMIC dataset in testing few samples T1-weighted MRI from IXI dataset. Scanner details are shown in Table~\ref{scanners}. ForkNet previously trained on the 17 subjects (all except subject case01020) from NAMIC dataset is used to test two subjects from IXI (Guys) datset (Figs.~\ref{IXIGuys025} and \ref{IXIGuys068}) and two subjects fro IXI (HH) dataset (Figs.~\ref{IXIHH518} and \ref{IXIHH519}). Images are processed using standard anisotropic scanner resolution that is clear in axial and coronal slices in Figs.~\ref{IXIGuys025} - \ref{IXIHH519}. The golden truth segmentation of IXI dataset is unavailable for detailed quantitative evaluation, nevertheless, it is still possible to validate potential usefulness visually. It is clear from these results that even with images obtained from different scanners (different resolution and magnet power), ForkNet can achieve a relatively good segmentation results especially for regions within brain and surrounding tissues. We observe some artifacts within the neck region, which can be obviously expected due to a absence of this region in the training dataset. These results encourage a further investigation of potential more general network architecture that can handle scanner variability.

%=================
% TABLE-6
%=================
\begin{table}
\centering
\footnotesize
\caption{Scanner details for original dataset used for training ForkNet and additional dataset used for robustness validation study.}
\label{scanners}
\begin{tabular}{ |l | c| c |c|}
\hline
{\bf Parameter} &{\bf Training dataset}	& \multicolumn{2}{c|}{\bf Testing dataset}  	 \\
\hline
\hline
Dataset & NAMIC  & IXI (Guys)& IXI (HH)\\
Scanner & GE &  Philips Intera& Philips Intera\\
%Location & Boston, Ma & London, UK \\
Magnet (T) & 3 & 1.5  & 3\\
Image (voxels) & 256$\times$256$\times$256 &  256$\times$256$\times$150 & 256$\times$256$\times$150\\
Resolution (mm) & 1$\times$1$\times$1 & 0.9375$\times$0.9375$\times$1.2  & 0.9375$\times$0.9375$\times$1.2 \\
\hline
\end{tabular}
\end{table}

%======================
% Figure-20
%======================

%======================
% Figure-21
%======================

%======================
% 5.4 TMS simulation
%======================

\subsection{TMS simulation}

The TMS-induced electric field computed using the network-generated head models successfully predicted the TMS-induced electric field computed using the original head model. In particular, the network prediction error was lower than 3\% in the targeted hand motor area. The error may be due to the CSF–GM segmentation in particular voxels. Moreover, the original model was constructed from T1 and T2 MRI images for the segmentation of brain and non-brain tissues \citep{Laakso2015BS}. The proposed method uses only T1 MRI images, which facilitates the setup in clinical practice for highly accurate head segmentation and TMS mapping. The construction of the head model of 1 mm resolution requires approximately 4.5 minutes, and the estimation of the induced electric field can be achieved in approximately 15 seconds. The relatively high computational cost for model construction is related to the large number of network variables. Future work is required to optimize the proposed network design to balance segmentation quality and computational cost.

%==========================
% 6. Conclusion
%==========================

\section{Conclusion}

A novel CNN architecture is proposed for the construction of human head models from MRI images. The proposed architecture (ForkNet) is different from conventional semantic CNNs because of the way the former handles individual decoders. This design offers several advantages that are difficult to evaluate in a single publication. Assigning individual decoder tracks for each anatomical structure could lead to wide-range network customization in terms of the convolutional kernel size, multi-resolution mapping, etc. ForkNet is used for head model construction through the segmentation of 13 different head anatomical tissues. The segmentation results obtained indicate the superior performance of the proposed method compared with other similar architectures, which are limited to brain segmentation only (in most cases). The network-developed head model was evaluated via TMS simulations, and our results indicate relatively high matching results between the induced electrical fields in the original model and the network-generated models. The total computation time (without GPU acceleration) required to conduct $n$ clinical TMS simulations for a single subject is approximately (4.5~+~0.25~$n$) minutes, which indicates a promising use in clinical applications.

Mathematica notebooks demonstrate the implementation of the ForkNet architecture, and trained networks are available for download at:

 \href{https://github.com/erashed/ForkNet}{https://github.com/erashed/ForkNet}.

%==========================
% Acknowledgment
%==========================

\section*{Acknowledgment}

This work was supported in part by the Ministry of Internal Affairs and Communications, Japan.

\bibliography{Refs}

%======================
% Figure-22 (SUPP)
%======================

\begin{figure*}
\centering
\includegraphics[width=1.2\textwidth]{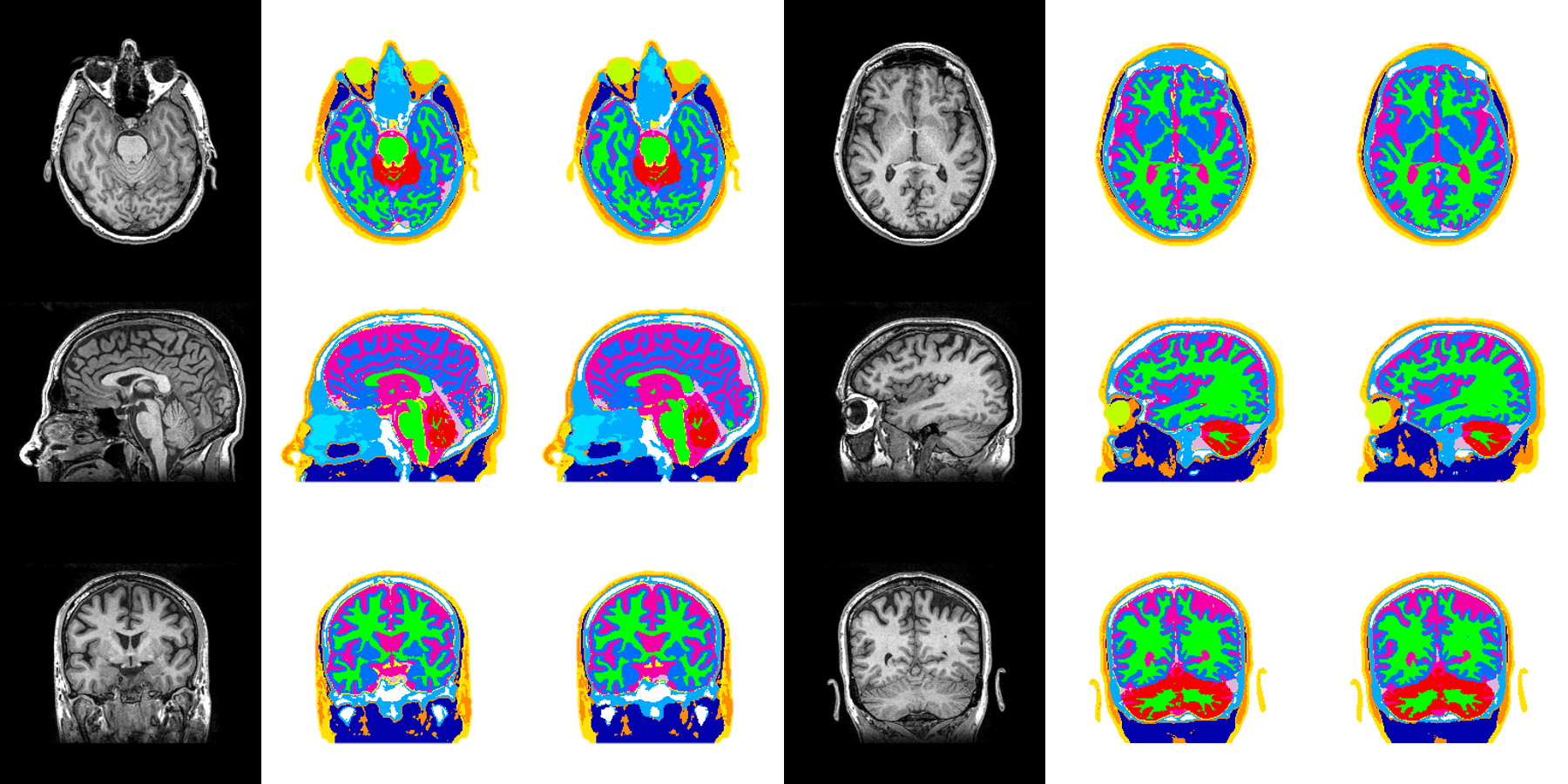}
\caption{Supplementary data for MRI (left), the original head model $R^\circ$ (middle), and the network-generated head model $R^\psi$ (right) for subject (case01015). From top to bottom, axial, sagittal, and coronal views are shown.}
\label{case01015}
\end{figure*}

%======================
% Figure-23 (SUPP)
%======================

\begin{figure*}
\centering
\includegraphics[width=1.2\textwidth]{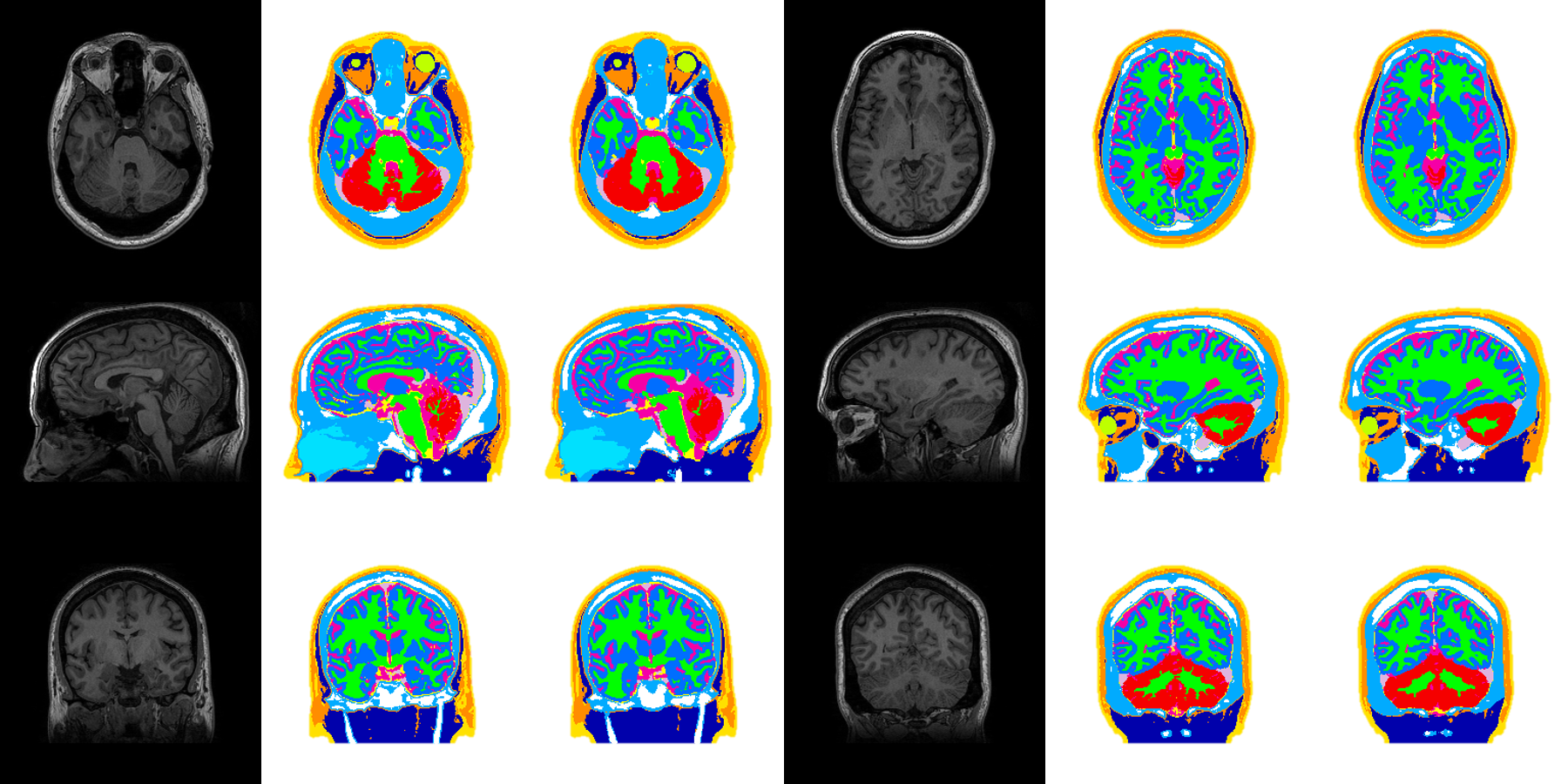}
\caption{Supplementary data for MRI (left), the original head model $R^\circ$ (middle), and the network-generated head model $R^\psi$ (right) for subject (case01017). From top to bottom,axial, sagittal, and coronal views are shown.}
\label{case01017}
\end{figure*}

%======================
% Figure-24 (SUPP)
%======================

\begin{figure*}
\centering
\includegraphics[width=1.2\textwidth]{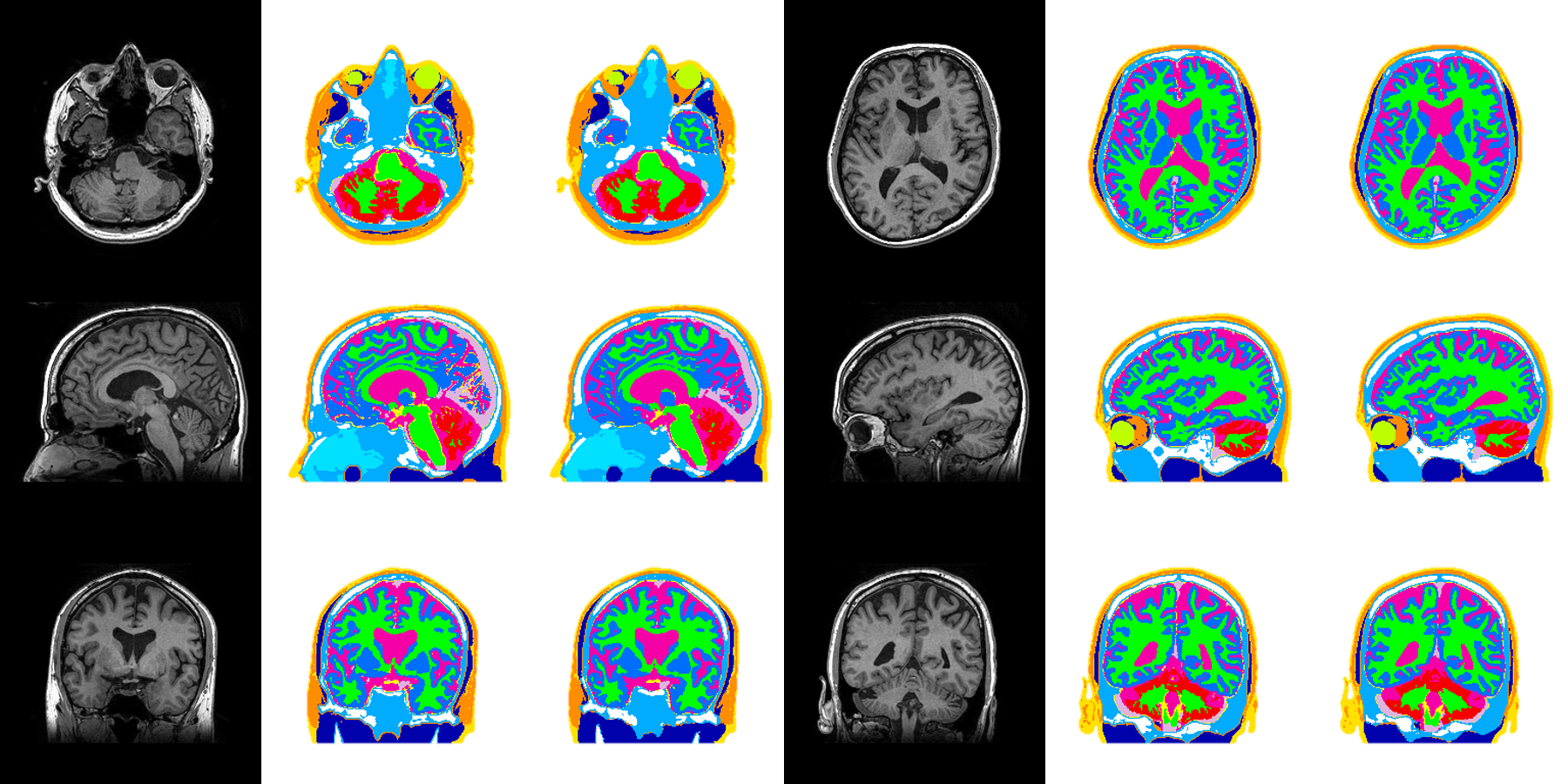}
\caption{Supplementary data for MRI (left), the original head model $R^\circ$ (middle), and the network-generated head model $R^\psi$ (right) for subject (case01018). From top to bottom, axial, sagittal, and coronal views are shown.}
\label{case01018}
\end{figure*}

%======================
% Figure-25 (SUPP)
%======================

\begin{figure*}
\centering
\includegraphics[width=1.2\textwidth]{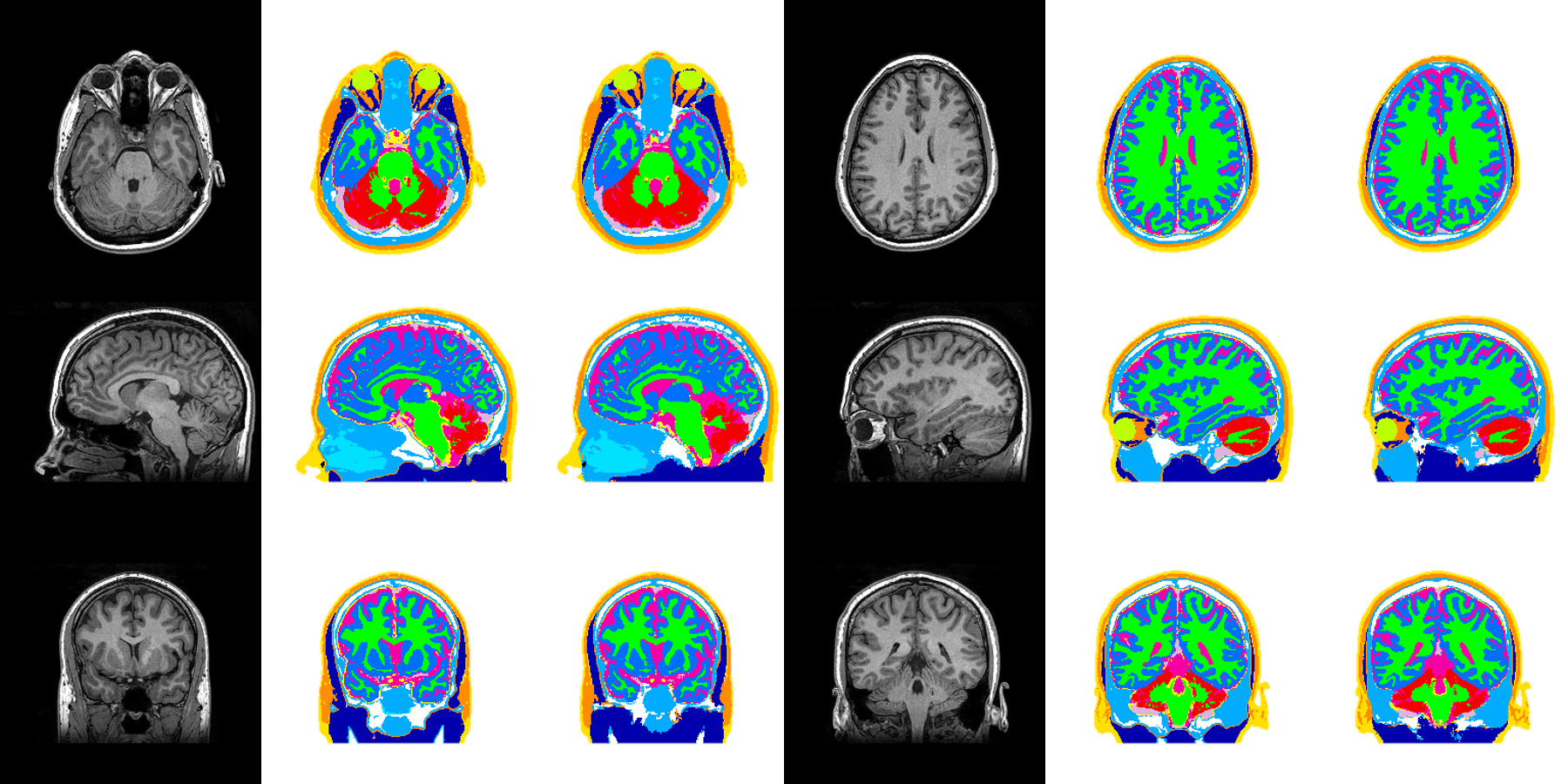}
\caption{Supplementary data for MRI (left), the original head model $R^\circ$ (middle), and the network-generated head model $R^\psi$ (right) for subject (case01025). From top to bottom, axial, sagittal, and coronal views are shown.}
\label{case01025}
\end{figure*}

%======================
% Figure-26 (SUPP)
%======================

\begin{figure*}
\centering
\includegraphics[width=1.2\textwidth]{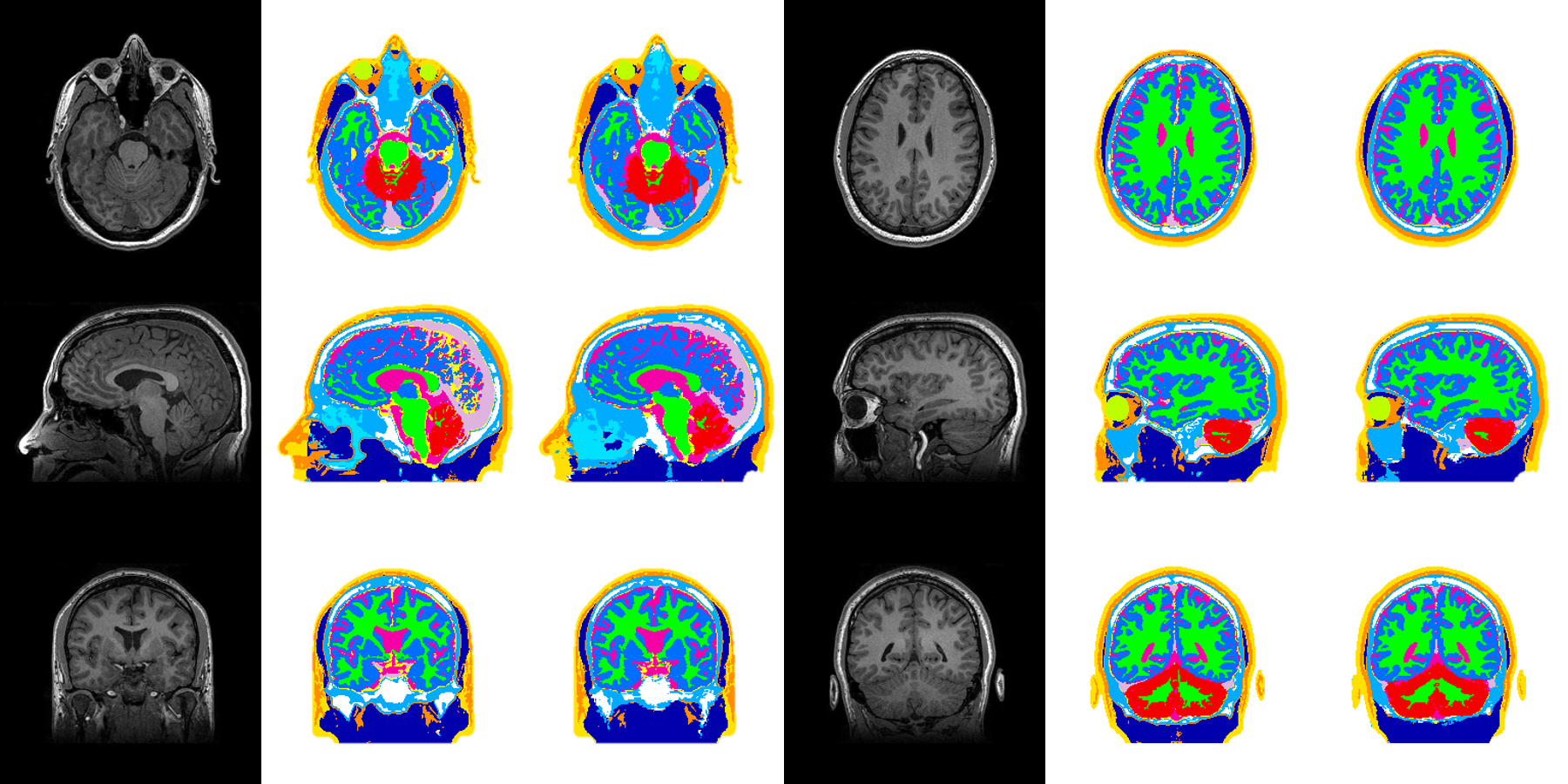}
\caption{Supplementary data for MRI (left), the original head model $R^\circ$ (middle), and the network-generated head model $R^\psi$ (right) for subject (case01026). From top to bottom, axial, sagittal, and coronal views are shown.}
\label{case01026}
\end{figure*}

%======================
% Figure-27 (SUPP)
%======================

\begin{figure*}
\centering
\includegraphics[width=1.2\textwidth]{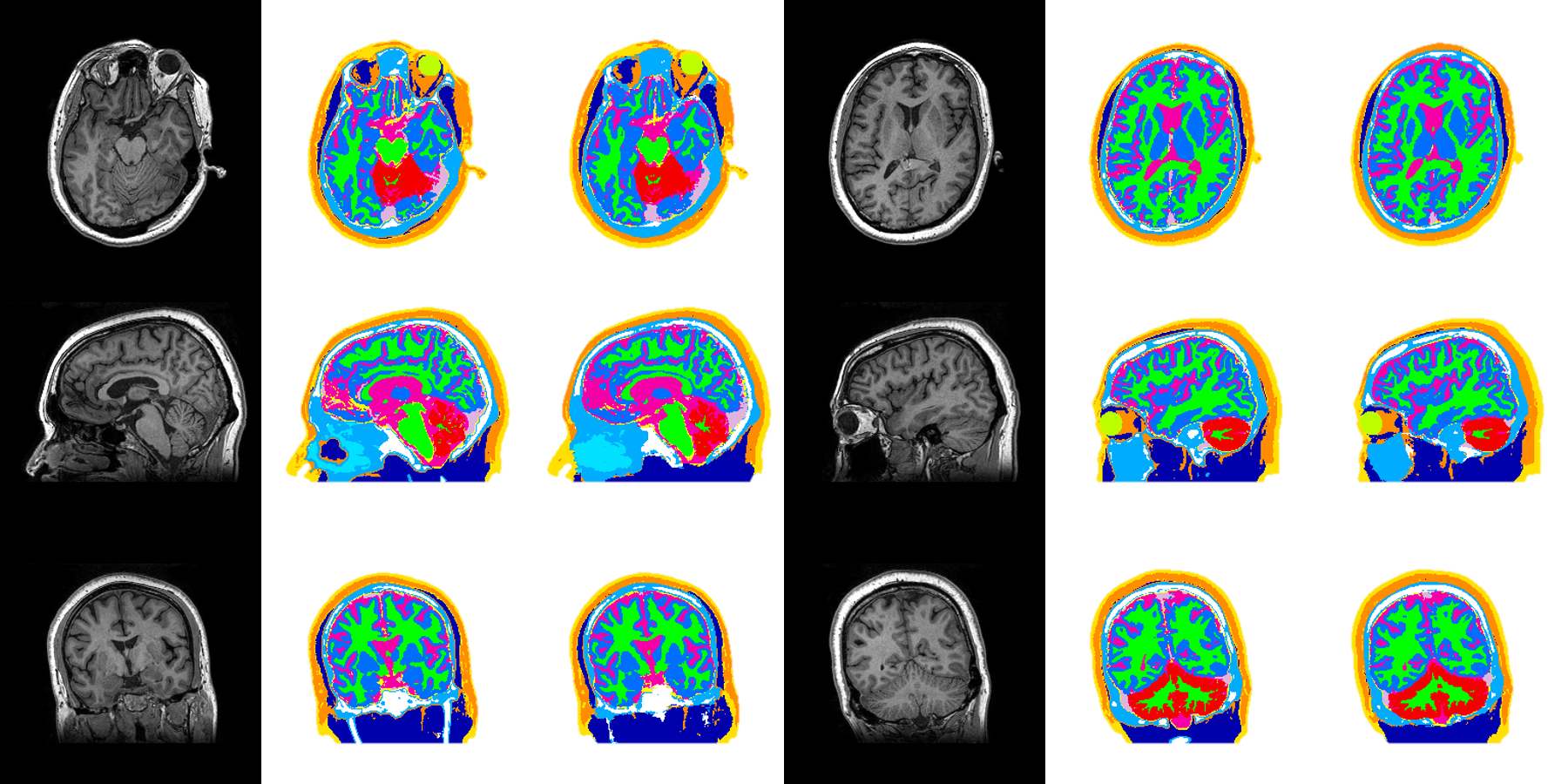}
\caption{Supplementary data for MRI (left), the original head model $R^\circ$ (middle), and the network-generated head model $R^\psi$ (right) for subject (case01029). From top to bottom, axial, sagittal, and coronal views are shown.}
\label{case01029}
\end{figure*}

%======================
% Figure-28 (SUPP)
%======================

\begin{figure*}
\centering
\includegraphics[width=1.2\textwidth]{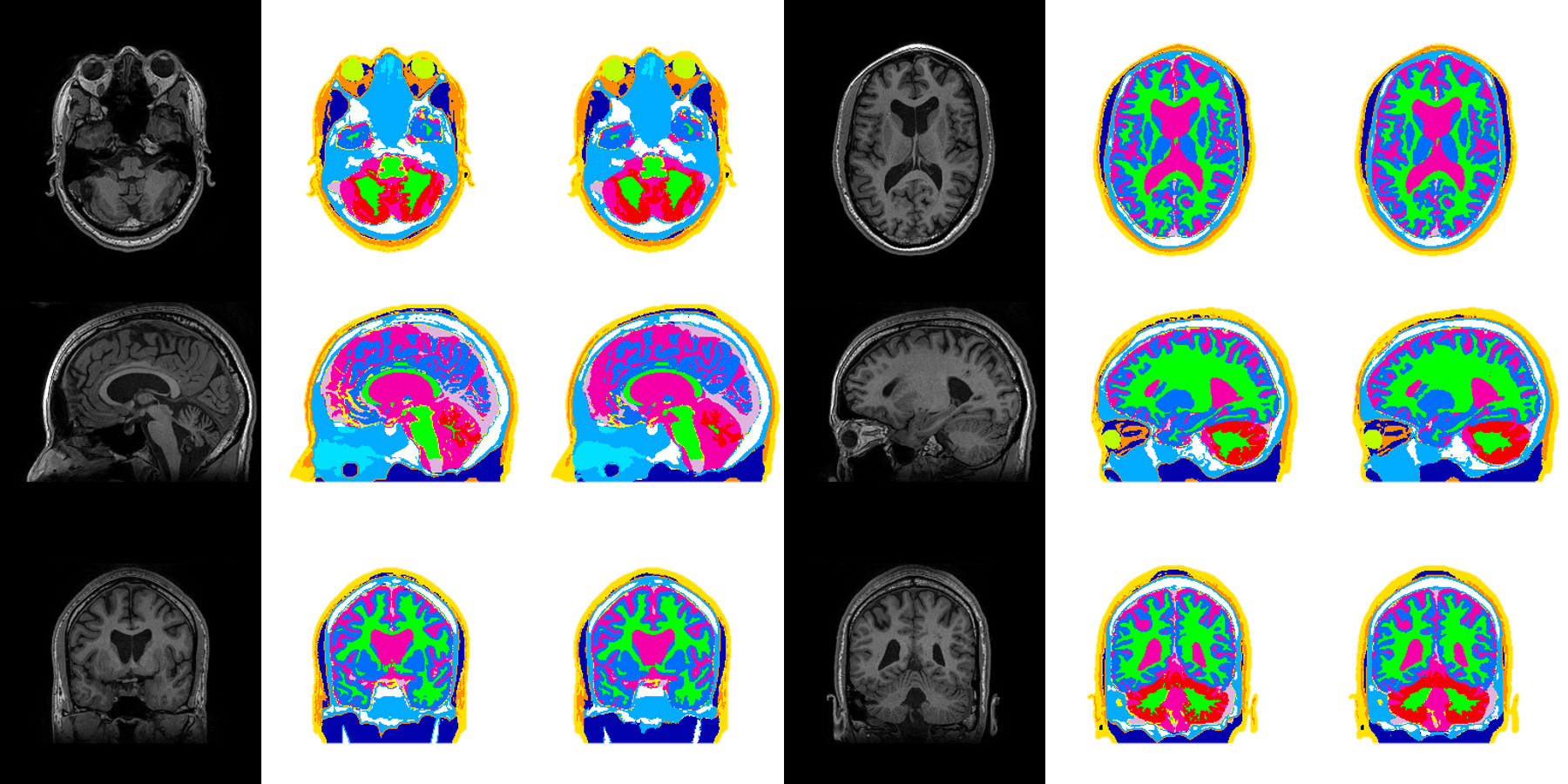}
\caption{Supplementary data for MRI (left), the original head model $R^\circ$ (middle), and the network-generated head model $R^\psi$ (right) for subject (case01034). From top to bottom, axial, sagittal, and coronal views are shown.}
\label{case01034}
\end{figure*}

%======================
% Figure-29 (SUPP)
%======================

\begin{figure*}
\centering
\includegraphics[width=1.2\textwidth]{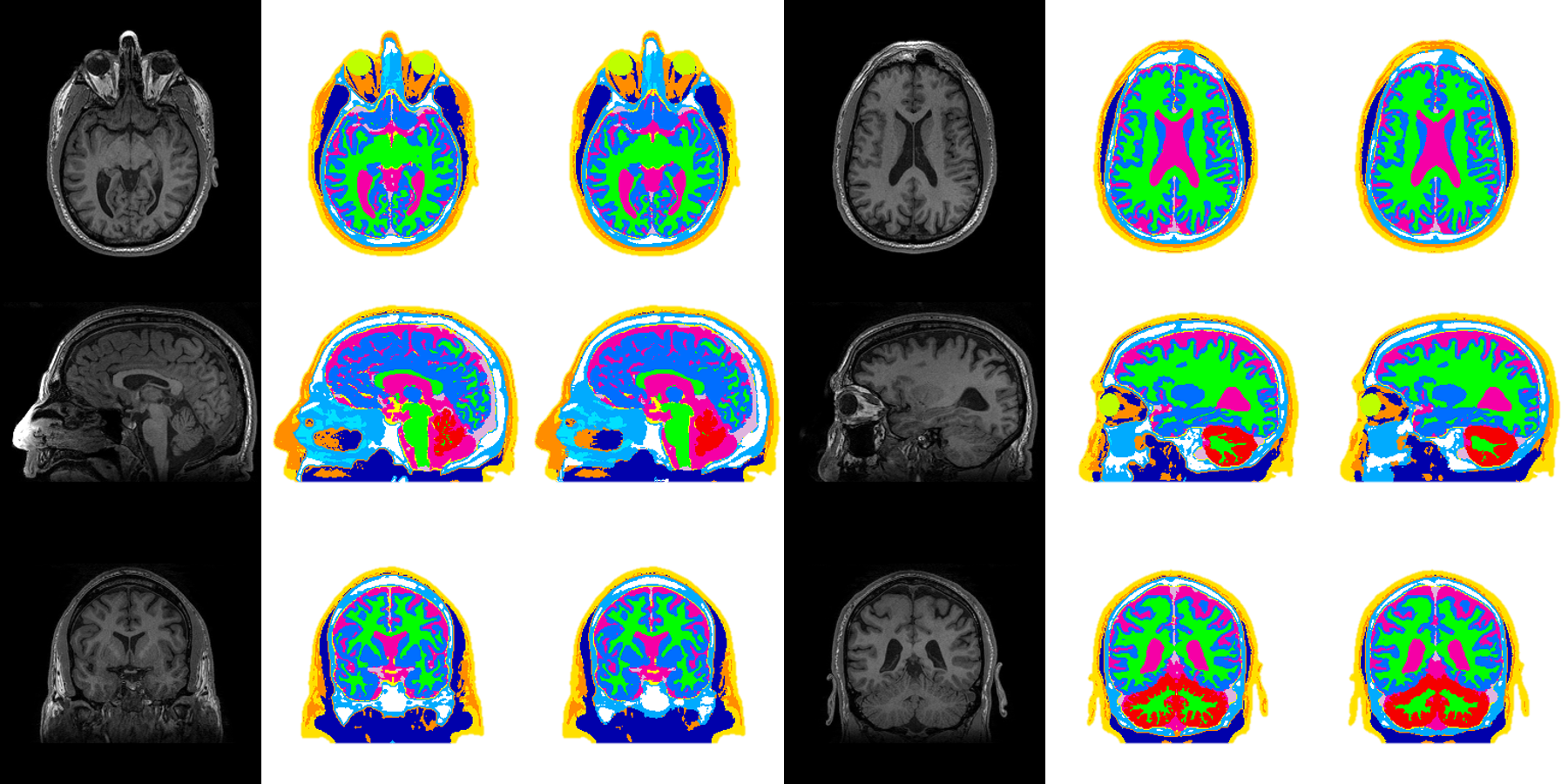}
\caption{Supplementary data for MRI (left), the original head model $R^\circ$ (middle), and the network-generated head model $R^\psi$ (right) for subject (case01045). From top to bottom, axial, sagittal, and coronal views are shown.}
\label{case01045}
\end{figure*}

%----------

\end{document}